\documentclass[twocolumn,aps,prl,superscriptaddress,amsmath,amssymb,floatfix,nofootinbib]{revtex4-2}
\usepackage[utf8]{inputenc}
\usepackage{color}
\usepackage{colortbl}
\usepackage{booktabs}
\usepackage{bm}
\usepackage{amsmath}
\usepackage{graphicx}
\usepackage[bookmarks=false,
breaklinks=false,pdfborder={0 0 1},backref=false,colorlinks=true]
{hyperref}
\hypersetup{
allcolors=blue}

\makeatletter

\providecommand{\tabularnewline}{\\}

\usepackage{bm}
\usepackage{physics}
\usepackage{array}
\usepackage{amsfonts}
\usepackage{esint}

\newcommand{\bk}{\mathbf{k}}

\newcommand{\br}{\mathbf{r}}

\newcommand{\ba}{\mathbf{a}}
\newcommand{\bA}{\mathbf{A}}
\newcommand{\bE}{\mathbf{E}}
\newcommand{\bB}{\mathbf{B}}
\newcommand{\intk}{\int_{\bk}}

\newcommand{\tcell}[2]{\begin{minipage}[t]{#1}\raggedright #2\end{minipage}}

\makeatother

\begin{document}
\title{Berry-Landau Fermi-liquid theory:  transport in presence of quantum geometry}
\author{Shuai A. Chen}
\email{chsh@pks.mpg.de}

\affiliation{Max Planck Institute for the Physics of Complex Systems, Nöthnitzer
    Straße 38, 01187 Dresden, Germany}
\author{Roderich Moessner}
\affiliation{Max Planck Institute for the Physics of Complex Systems, Nöthnitzer
    Straße 38, 01187 Dresden, Germany}
\date{\today}

\begin{abstract}
    Landau Fermi-liquid theory characterizes interacting metals through quasiparticles and their residual interactions. It is a challenge to incorporate non-trivial quantum geometry -- as encoded by Berry phases and the quantum metric -- as a fundamental ingredient.
    We formulate a Berry-Landau Fermi-liquid theory for spinless fermions within an
    isolated band crossing the Fermi surface and derive the Landau functional
    through the Nozières--Luttinger construction. The instantaneous response to a
    perturbation of the Bloch waves generates both an anomalous Berry-connection
    potential in the action and an interaction-induced quantum-geometric
    contribution to the quasiparticle current. The conserved physical charge
    current is then obtained via the electromagnetic Ward identity as a combination
    of this quantum-geometric current and the bare drift current. Therefore, the
    intrinsic anomalous Hall conductivity is fixed by the Berry-curvature integral
    of the dressed quasiparticle band while the Drude weight contains both
    conventional and quantum-geometric contributions. In the presence of Galilean
    symmetry, the Drude weight is protected against interaction renormalization. In
    the flat/narrow-band limit, transport is dominantly quantum-geometric and can
    be thermally enhanced. These results establish quasiparticle occupations,
    Landau interactions, and the quantum geometry carried by quasiparticles as the
    fundamental low-energy ingredients of a Berry-Landau Fermi liquid.
\end{abstract}
\maketitle
\makeatletter\let\arxiv@addcontentsline\addcontentsline
\renewcommand{\addcontentsline}[3]{}\makeatother

\emph{\color{blue} Introduction.---} Landau's Fermi-liquid (FL)
theory describes an interacting metal through long-lived quasiparticles
labeled by momentum, with the low-energy state specified by their
occupations. In the presence of non-trivial quantum geometry (QG), this needs to be supplemented by information about the Berry phase and Hilbert-space geometry \cite{MarzariVanderbilt1997,Berry1984,KarplusLuttinger1954,TKNN1982,ProvostVallee1980,Resta2011,SouzaWilkensMartin2000,Haldane2004}.
Far from passive, this geometry controls transport---the anomalous
velocity, orbital magnetization, linear and nonlinear Hall responses,
and positional shifts of the carriers \cite{Xiao2010,XiaoShiNiu2005,Mitscherling2020,MitscherlingHolder2022,SodemannFu2015,Ulrich2026,GaoXiao2019,Gao2023,Wang2023,AhnGuoNagaosaVishwanath2022,GaoYangNiu2014}---and
enters the interaction itself through the form factors of band projection,
driving, e.g., flat-band superconductivity governed by the quantum
metric \cite{MitscherlingAvdoshkinMoore2025,PeottaTorma2015,JulkuEtAl2016,Liang2017,Torma2023,xkxw-1134,TormaPeottaBernevig2022,ChenLaw2024,rw6g-w7my,zdyq-3m9x,2023PhRvL.130v6001H}.
Quantum-geometric transport is realized in topological semimetals \cite{ArmitageMeleVishwanath2018,Nakatsuji2015,LiuCoSnS2018},
kagome and pyrochlore metals \cite{WilsonOrtiz2024,Ortiz2020,Ye2018},
and moiré materials \cite{BistritzerMacDonald2011,2025npjQM..10..101Y,2020NatPh..16..725B},
and QG is now measurable \cite{2025NatPh..21..110K,Kim2025QuantumMetric}. The question thus arises of how QG is most naturally incorporated into FL theory.

The notion that it is the Fermi surface which determines transport properties
extends to topological transport phenomena: the Berry connection on the Fermi
surface determines the intrinsic anomalous Hall effect (AHE)
\cite{SundaramNiu1999,Nagaosa2010,Xiao2010,Haldane2004,PhysRevB.76.195109}.
Berry-curvature effects enter kinetic theory at the single-particle level
\cite{StephanovYin2012,ParkBalents2026} and have been combined with
interaction-dressed vertices in FL transport
\cite{ShindouBalents2006,SonYamamoto2012,ChenSon2017,2024PhRvB.109w5146H,PasquaFabrizio2025}.
In these approaches, QG is imported from the underlying multiband structure or
postulated in a phenomenological action, rather than constructed from just the
Fermi surface itself. Recently, we proposed a generalized Peierls substitution
\cite{xkxw-1134}. There, we noted that the current density operator acquires a
QG contribution in the presence of interactions. In this work, we show how to
incorporate into FL theory such QG effects, alongside the standard ingredients
of quasiparticle occupations and Landau parameters.

Towards a microscopic formulation of FL theory, the Nozières--Luttinger
construction establishes that the Landau functions encode the residual forward
scattering between quasiparticles
\cite{NozieresLuttinger1962,LuttingerNozieres1962} through the particle-hole
Bethe-Salpeter equation \cite{Landau1957,Silin1958,Nozieres,BaymPethick1991}.
Adapting this construction, we develop a Berry-Landau-FL (BL-FL) theory by projecting an
interacting multiorbital lattice model onto an isolated band, and extract
physical consequences such as the resulting transport properties.
Finally, we verify these predictions using Hartree-Fock calculations for an
isolated Chern band of the Wilson-Dirac model
\cite{Wilson1974,QiWuZhang2006,Shen2012}.

\emph{\color{blue} System setup.---}
To treat quantum-geometric effects in an interacting metal, we
start with a generic $N_\mathrm{orb}$-orbital lattice Hamiltonian $H_{0}=\sum_{\mathbf{k}}H_{0}(\mathbf{k}) $ with
\begin{equation}
    H_{0}(\mathbf{k})=\sum^{N_{{\rm orb}}}_{\alpha,\beta=1}t_{\alpha\beta}(\mathbf{k})c^{\dagger}_{\mathbf{k}\alpha}c_{\mathbf{k}\beta}^{},
\end{equation}
where $c_{\mathbf{k}\alpha}$ annihilates a fermion with crystal momentum
$\mathbf{k}$ and orbital index $\alpha$. Among the $N_\mathrm{orb}$ bands, we assume that $H_{0}$
hosts an isolated (active) band crossing the Fermi energy, separated from all
remote bands by an energy gap $\Delta$. Throughout this work, we take the  isolated-band limit, $\Delta\to\infty$,
at fixed active-band width, interaction strength, and temperature.
We allow for the active band to be topologically nontrivial.
We thus project onto the Hilbert subspace $\mathbb{H}^{{\rm sub}}$ of the active band. A general short-ranged interaction is
considered for the development of the theory.

\emph{\color{blue} Anomalous potential from band projection without interactions.---}
Consider first the noninteracting problem. The Hilbert subspace $\mathbb H^\mathrm{sub}$ is spanned by the active band at each crystal momentum $\mathbf{k}$, $H_{0}(\mathbf{k})|u_{\mathbf{k}}\rangle=\epsilon^{0}_{\mathbf{k}}|u_{\mathbf{k}}\rangle$.
Then the projection onto $\mathbb H^\mathrm{sub}$ can  be imposed
via a Grassmann-valued Lagrange multiplier $\lambda_{\mathbf{k}\alpha}$~\cite{xkxw-1134},
\begin{equation}
    S_{\lambda}=\int dt\sum_{\mathbf{k}\alpha}\left[\lambda^{\dagger}_{\mathbf{k}\alpha}\left(c_{\mathbf{k}\alpha}-u_{\mathbf{k}\alpha}f_{\mathbf{k}}\right)+\mathrm{h.c.}\right],\label{eq:Slambda}
\end{equation}
where the operator $f_{\mathbf{k}}$ annihilates a fermion at crystal momentum $\mathbf{k}$ of the active band.
We have the partition function  $Z=\int DcDc^\dagger DfDf^\dagger D\lambda D\lambda^\dagger \,e^{i(S_{0}+S_{\lambda})}$ with the action
$S_{0}=\int
    dt\sum_{\mathbf{k}}[\sum_{\alpha}c^{\dagger}_{\mathbf{k}\alpha}i\partial_{t}c_{\mathbf{k}\alpha}-H_{0}(\mathbf{k})]$.
In the absence of external perturbations, integrating out $\lambda$ yields the
free theory, $S[f]=\int
    dt\sum_{\mathbf{k}}f^{\dagger}_{\mathbf{k}}(i\partial_{t}-\epsilon^{0}_{\mathbf{k}}+\mu)f_{\mathbf{k}}$.

QG manifests itself through the response to external perturbations. Taking a
uniform electric field as an example, the Bloch waves become explicitly time
dependent, and this time dependence encodes the Berry connection. We choose the
gauge $\mathbf{A}(t)=-\mathbf{E}t$, such that
$H_{0}(\mathbf{k},t)=H_{0}(\mathbf{k}+e\mathbf{A}(t))$. In the weak-field
limit, the instantaneous Bloch wave follows the Houston
trajectory~\cite{Houston1940,PhysRev.117.432},
$|u_{\mathbf{k}}(t)\rangle=|u_{\mathbf{k}(t)}\rangle$ with
$\dot{\mathbf{k}}=-e\mathbf{E}$. Accordingly, the projection in $S_\lambda$ is
imposed in the moving Bloch frame,
$c_{\mathbf{k}\alpha}=u_{\mathbf{k}(t)\alpha}f_{\mathbf{k}}$. After projection,
the time-derivative term produces an additional quantum-geometric contribution,
\begin{equation}
    \sum_{\alpha}c^{\dagger}_{\mathbf{k}\alpha}i\partial_{t}c^ {}_{\mathbf{k}\alpha}\to f^{\dagger}_{\mathbf{k}}i\partial_{t}f^{}_{\mathbf{k}}+e\mathbf{E}\cdot\mathbf{a}(\mathbf{k})\,f^{\dagger}_{\mathbf{k}}f^ {}_{\mathbf{k}},\label{eq:timeproj}
\end{equation}
with the Berry connection $\mathbf{a}(\mathbf{k})=-i\langle u_{\mathbf{k}}|\nabla_{\mathbf{k}}u_{\mathbf{k}}\rangle$.
The nonzero value of $\dot{\mathbf k}$ is crucial for the appearance of the second term. 
For an electric field, it gives the Berry-connection contribution to the Wannier--Stark energy shift \cite{1959JPCS...10..286A}. 
We refer to it as the anomalous potential because it generates the topological response, as we show later. The appearance of the anomalous potential is essential for gauge invariance due to the emergent gauge structure of the band projection~\cite{gauge_note}.

The anomalous potential generates the intrinsic AHE,
\begin{align}
    \sigma_{xy}
     & =\frac{e^{2}}{\hbar}\int\frac{d^{2}\mathbf{k}}{(2\pi)^{2}}
    \left[-\partial_{\epsilon}n_F(\xi_{\mathbf{k}})\right]
    \left[v_{\mathbf{k} x}^0a_y(\mathbf{k})
        -v_{\mathbf{k}y}^0 a_x(\mathbf{k})\right]
    \notag                                                        \\
     & =\frac{e^{2}}{\hbar}\int\frac{d^{2}\mathbf{k}}{(2\pi)^{2}}
    n_F(\xi_{\mathbf{k}})\Omega_{xy}(\mathbf{k}),
    \label{eq:sigmaxy-kubo}
\end{align}
where $\xi_{\mathbf{k}}=\epsilon^0_{\mathbf{k}}-\mu$ and
$\Omega_{xy}=\partial_{k_x}a_y-\partial_{k_y}a_x$.
The first line makes explicit the Fermi-surface character of the metallic AHE,
consistent with Haldane's formulation \cite{Haldane2004}.
To see its origin, for a spatially uniform vector
potential with $\mathbf{E}(\omega)=i\omega\mathbf{A}(\omega)$, integrating out
$f$ gives
$
    S_{\mathrm{eff}}[A]
    =-i\,\mathrm{Tr}\ln\!\left(G^{-1}_{f0}-V[A]\right),
$
where $
    G^{-1}_{f0}
    =\omega-\epsilon^0_{\mathbf{k}}+i0^+,
$ and $
    V[A]
    =e\mathbf{A}\cdot\mathbf{v}^0
    -ie\omega\,\mathbf{A}\cdot\mathbf{a}$.
This defines the band and Berry-connection current operators,
$
    \hat {\mathbf J}^{(\mathrm{v})}
    =
    e\sum_{\mathbf{k}}
    \mathbf v^0_{\mathbf{k}}
    f^\dagger_{\mathbf{k}}f_{\mathbf{k}}^{},
$ and $
    \hat {\mathbf J}^{(\mathrm{B})}
    =
    -ie\omega\sum_{\mathbf{k}}
    \mathbf{a}(\mathbf{k})
    f^\dagger_{\mathbf{k}}f_{\mathbf{k}}^{}
$.
The intrinsic AHE arises from the current-current correlation between
$\hat{\mathbf J}^{(\mathrm{v})}$ and
$\hat{\mathbf J}^{(\mathrm{B})}$ in the Kubo response.
More generally, any slow perturbation exerting a generalized force
$\dot{\mathbf{k}}$---a magnetic field, a smooth potential, a temperature
gradient, strain, or a frame rotation---generates the anomalous potential
$\dot{\mathbf{k}}\cdot\mathbf{a}(\mathbf{k})\,f^{\dagger}_{\mathbf{k}}f_{\mathbf{k}}^{}$. The corresponding current operators and representative responses are
summarized in Table~\ref{tab} in the End
Matter.

\emph{\color{blue}Adding interactions and the QG current to Landau theory.---}
Interactions play two roles: they modify the band structure and they generate an interaction-induced QG current \cite{xkxw-1134}. The former determines the active Hilbert subspace. To capture this, we promote the Bloch waves $u_{\mathbf{k}\alpha}$ to variational parameters in Eq.~\eqref{eq:Slambda}. Stationarity identifies the retained state with the on-shell quasiparticle state,
\begin{equation}
    G^{-1}(\omega,\mathbf{k})|u_{\mathbf{k}}\rangle=0,
\end{equation}
at the quasiparticle pole of the full Green function in the full Hilbert space. At $T=0$, this construction is unambiguous on the Fermi surface, where the quasiparticle decay rate vanishes and the Hermitian matrix $G^{-1}(0,\mathbf{k}_F)$ has $|u_{\mathbf{k}_F}\rangle$ as its zero mode \cite{Haldane2004,IshikawaMatsuyama1986ZPC,IshikawaMatsuyama1987NPB}.
In the $\Delta\to\infty$ limit, the hybridization with remote bands is
eliminated and the low-energy dynamics thus remains confined to this projected
subspace.

Regarding the latter, interactions projected onto the subspace acquire form
factors, whose dependence on an added perturbation as described above
generates a contribution to the current operator.
Before many-body renormalization, the projected interaction in the
forward-scattering channel takes the form $H_\mathrm{int}\rightarrow
    V^{-1}\sum_{\mathbf k\mathbf k^\prime} f^0_{\mathbf k\mathbf k^\prime}
    n_{\mathbf k}n_{\mathbf k^\prime}$, with $n_{\mathbf k}=f_{\mathbf k}^\dagger
    f_{\mathbf k}$. As the bare forward-scattering function $f^0_{\mathbf k\mathbf
    k^\prime}$ contains the form factor, its response to a uniform vector potential
gives the bare QG current operator $\hat J^{(\rm QG,0)}_{i} = \frac{e}{2V}
    \sum_{\mathbf{k}\mathbf{k}'} \left( \nabla_{\mathbf{k}} + \nabla_{\mathbf{k}'}
    \right)_i f^{(0)}_{\mathbf{k}\mathbf{k}'} n_{\mathbf{k}}n_{\mathbf{k}'}$.
Many-body processes then renormalize the bare forward-scattering function $
    f^{0}_{\mathbf{k}\mathbf{k}'} $, which is the tree-level contribution to the
amputated connected four-point vertex $\Gamma$, into the Landau function
\begin{equation}
    f_{\mathbf{k}\mathbf{k}'}
    =
    Z_{\mathbf{k}}Z_{\mathbf{k}'}
    \lim_{\omega\to0}\lim_{\mathbf q\to0}
    \Gamma(\mathbf{k},\mathbf{k}';i\omega,\mathbf q),
\end{equation}
where the dynamic limit is defined by taking the  $\mathbf q\to0$ limit before $\omega\to0$.
Correspondingly, this renormalized interaction generates the interaction correction to the bare QG current operatotr $\hat J^{(\rm QG,0)}$, yielding
the renormalized QG current operator in the dynamic limit,
\begin{equation}
    \hat {\mathbf J}^{(\rm QG,\omega)} = \frac{e}{2V}
    \sum_{\mathbf{k}\mathbf{k}'} \left( \nabla_{\mathbf{k}} + \nabla_{\mathbf{k}'}
    \right) f_{\mathbf{k}\mathbf{k}'} n_{\mathbf{k}}n_{\mathbf{k}'}.
    \label{eq:JQGomega}
\end{equation}
This provides the  foundation to construct the energy functional in the BL-FL theory.

\emph{\color{blue} Energy functional.---}
These results assemble into the BL-FL functional,
$
    \delta E_{\mathrm{BL\text{-}FL}}[n,\mathbf A]
    =
    \delta E_{\mathrm L}[n]
    +
    \delta E_{\mathrm{ext}}[n,\mathbf A].˝
$
In the absence of explicit coupling to external perturbations, the first part retains the familiar Landau form,
\begin{equation}
    \delta E_{\mathrm L}[n]
    =
    \sum_{\mathbf{k}}
    \epsilon^{\mathrm{qp}}_{\mathbf{k}}
    \delta n_{\mathbf{k}}
    +
    \frac{1}{2V}
    \sum_{\mathbf{k}\mathbf{k}'}
    f_{\mathbf{k}\mathbf{k}'}
    \delta n_{\mathbf{k}}\delta n_{\mathbf{k}'},
    \label{eq:functional-main}
\end{equation}
where $\delta n_{\mathbf{k}}=n_{\mathbf{k}}-n^0_{\mathbf{k}}$ and
$\mathbf v^{\mathrm{qp}}_{\mathbf{k}}
    =\nabla_{\mathbf{k}}\epsilon^{\mathrm{qp}}_{\mathbf{k}}$. Here, a perturbation enters via a change in the occupation numbers.

The second term, upon coupling to external perturbations, is obtained by
expanding around the equilibrium distribution (and dropping field-dependent
constants):
\begin{align}
    \delta E_{\mathrm{ext}}[n,\mathbf A]
    = &
    \sum_{\mathbf{k}}
    \left[
        \mathbf A\cdot\bm{\mathcal J}_{L}(\mathbf{k})
        +
        \dot{\mathbf{k}}\cdot\mathbf a(\mathbf{k})
        \right]\delta n_{\mathbf{k}}
    \notag \\
      & +
    \frac{1}{2V}
    \sum_{\mathbf{k}\mathbf{k}'}
    \mathbf A\cdot\bm{\Lambda}_{\mathbf{k}\mathbf{k}'}
    \delta n_{\mathbf{k}}\delta n_{\mathbf{k}'},
    \label{eq:Lambda-expanded}
\end{align}
with the quasiparticle drift current $ e\mathbf v^{\mathrm{qp}}_{\mathbf{k}}$,
the longitudinal current
$
    \bm{\mathcal J}_{L}(\mathbf{k})
    \equiv
    e\left[
    \mathbf v^{\mathrm{qp}}_{\mathbf{k}}
    +
    \frac{1}{V}\sum_{\mathbf{k}'}
    n^0_{\mathbf{k}'}
    \nabla_{\mathbf{k}'}
    f_{\mathbf{k}\mathbf{k}'}
    \right]
$,
and the QG kernel
$ \bm{\Lambda}_{\mathbf{k}\mathbf{k}'}
    \equiv
    e\left(
    \nabla_{\mathbf{k}}
    +
    \nabla_{\mathbf{k}'}
    \right)
    f_{\mathbf{k}\mathbf{k}'}$.
Eq.~\eqref{eq:Lambda-expanded} displays the three quantum-geometric structures of the BL-FL. The first is the dressed longitudinal current, including Landau backflow; the second is the anomalous Berry potential associated with the time dependence of the Bloch waves, with $\dot{\mathbf k}$ determined by the semiclassical force of the external perturbation; and the last is a bilinear QG coupling between occupation fluctuations. For an electromagnetic field,
$
    \dot{\mathbf{k}}
    =
    -e\left(
    \mathbf E
    +
    \mathbf v^{\mathrm{qp}}_{\mathbf{k}}\times\mathbf B
    \right).
$
The connection to the QG current operator in Eq.~\eqref{eq:JQGomega} is transparent.
The coupling term between the gauge field and $\hat{\mathbf J}^{(\rm QG,\omega)}$
is mapped as $ \mathbf A\cdot\hat{\mathbf J}^{(\rm QG,\omega)} \rightarrow
    \frac{1}{2V} \sum_{\mathbf{k}\mathbf{k}'} \mathbf
    A\cdot\bm{\Lambda}_{\mathbf{k}\mathbf{k}'} n_{\mathbf{k}}n_{\mathbf{k}'} . $
Expanding \(n_{\mathbf{k}}=n^0_{\mathbf{k}}+\delta n_{\mathbf{k}}\), the term
linear in \(\delta n_{\mathbf{k}}\) gives the interaction-induced
single-quasiparticle current, $ \bm{\gamma}^{(\mathrm{int}),\omega}(\mathbf{k})
    \equiv \frac{1}{V} \sum_{\mathbf{k}'} \bm{\Lambda}_{\mathbf{k}\mathbf{k}'}
    n^0_{\mathbf{k}'}, $ so that $ \bm{\mathcal J}_{L}(\mathbf{k}) = e\mathbf
    v^0_{\mathbf{k}} + \bm{\gamma}^{(\mathrm{int}),\omega}(\mathbf{k}). $ The two
derivatives in \(\bm{\Lambda}_{\mathbf{k}\mathbf{k}'}
=e(\nabla_{\mathbf{k}}+\nabla_{\mathbf{k}'}) f_{\mathbf{k}\mathbf{k}'}\) have
the familiar Landau interpretation: the \(\nabla_{\mathbf{k}}\) term gives the
quasiparticle velocity shift \(e(\mathbf v^{\mathrm{qp}}-\mathbf v^0)\), while
the \(\nabla_{\mathbf{k}'}\) term takes the Landau-backflow form after
integration by parts over the Brillouin zone. The term quadratic in \(\delta
n_{\mathbf{k}}\delta n_{\mathbf{k}'}\) is the two-quasiparticle part of
the QG current and contributes to nonlinear response and current fluctuations.

Charge conservation implies a Ward identity which fixes the longitudinal current $\bm{\mathcal J}_{L}$ in relation to the
bare current $e\mathbf v^0_{\mathbf{k}}$,
\begin{equation}
    \bm{\mathcal J}_L(\mathbf{k})
    =
    e\mathbf v^0_{\mathbf{k}}
    +
    \bm{\gamma}^{(\mathrm{int}),\omega}(\mathbf{k}).
    \label{eq:ward-main}
\end{equation}
Thus, $\bm{\mathcal J}_{L}$ is the conserved physical current carried by a quasiparticle, including the interaction-induced velocity renormalization and Landau backflow.
The Berry-connection current is transverse and is
not constrained by this longitudinal identity.
When Galilean invariance emerges in the long-wavelength limit, 
 the Landau function depends only on the momentum transfer,
$f_{\mathbf{k}\mathbf{k}'}=f(\mathbf{k}-\mathbf{k}')$, and thus
$
    \left(
    \nabla_{\mathbf{k}}
    +
    \nabla_{\mathbf{k}'}
    \right)
    f_{\mathbf{k}\mathbf{k}'}
    =0.
$ Therefore, the QG contribution vanishes and the physical current reduces to
\(\bm{\mathcal J}_{L}=e\mathbf v^0\), recovering the conventional Landau
relation.

The uniform collisionless response then follows from  the free energy $F[n,\mathbf A]= \delta E_{\mathrm{BL\text{-}FL}}-TS[n]$ with the entropy $S[n]= - \sum_\mathbf{k}[n_\mathbf{k}\ln n_\mathbf{k}+ (1-n_\mathbf{k})\ln (1-n_\mathbf{k}) ]$. 
Concretely, we can obtain $F[n,\mathbf A]$ by substituting the linear-order quasiparticle energy variation $
    \delta\epsilon^{\mathrm{qp}}_{\mathbf{k}} = \mathbf A\cdot\bm{\mathcal
        J}_{L}(\mathbf{k}) + \dot{\mathbf{k}}\cdot\mathbf a(\mathbf{k}) +
    \frac{1}{V}\sum_{\mathbf{k}'} f_{\mathbf{k}\mathbf{k}'} \delta n_{\mathbf{k}'},
$ with $ \delta n_{\mathbf{k}} = \frac{\partial n^0_{\mathbf{k}}}
    {\partial\epsilon^{\mathrm{qp}}_{\mathbf{k}}}
    \delta\epsilon^{\mathrm{qp}}_{\mathbf{k}}$. 
Thus the longitudinal response is
governed by $\bm{\mathcal J}_{L}$, whereas the anomalous
potential supplies the transverse quantum-geometric response.
In particular, the Drude weight
and intrinsic AHE follow, respectively, from the longitudinal $\mathbf{A}$--$\mathbf{A}$ and
antisymmetric $\mathbf A$--$\mathbf E$ responses obtained from the free energy $ F[n,\mathbf A]$.

\emph{\color{blue} AHE and transverse Wiedemann-Franz law.---}
The electric field produces the band-projected action term 
$e\mathbf E\cdot\mathbf a$, which is the
Berry-connection contribution to the Wannier--Stark energy shift.
Then differentiating the free energy $F[n,\mathbf A]$
with respect to $\mathbf A$ gives the physical current.
Thus, we obtain the AHE response 
        $\sigma^{\mathrm{AHE}}_{xy} =
    \frac{e}{V}\sum_{\mathbf k\mathbf k^\prime}\left\{ \mathcal J_{L,x}(\mathbf k) [(1+\chi f)^{-1}\chi]_{\mathbf k\mathbf k^\prime}  a_y(\mathbf k^\prime)
        -(x\leftrightarrow y) \right\}$  where $\chi$ denotes the quasiparticle susceptibility with  $\chi_{\mathbf k\mathbf k^\prime}\equiv - \frac{\partial n_F}
    {\partial\epsilon^{\mathrm{qp}}_{\mathbf{k}}}\delta_{\mathbf k \mathbf k^\prime}$ and $f$ denotes the Landau function $f_{\mathbf k\mathbf k^\prime}$. 
The Ward identity in Eq.~\eqref{eq:ward-main}
then gives the AHE as a Fermi-surface integral
\begin{equation}
    \sigma_{xy}
    =
    \frac{e^2}{V}\sum_{\mathbf{k}}\chi_{\mathbf{k}}
    \left[
        v^\mathrm{qp}_{\mathbf{k}x}a_y(\mathbf{k})
        -
        v^\mathrm{qp}_{\mathbf{k}y}a_x(\mathbf{k})
        \right],
\end{equation} with
$
    \chi_{\mathbf{k}}
    =
    -\frac{\partial n_F}
    {\partial\epsilon^{\mathrm{qp}}_{\mathbf{k}}}.
$
Thus, the Landau-backflow dressed current $\mathcal J_L(\mathbf k)=\sum_{\mathbf k^\prime }(1+\chi f)_{\mathbf k\mathbf k^\prime}\mathbf v^\mathrm{qp}_\mathbf{k}$ cancels exactly against  the Landau-backflow dressed Berry connection 
$\bm{\mathcal A}(\mathbf k) = \sum_{\mathbf k^\prime}[(1+f\chi)^{-1}]_{\mathbf k\mathbf k^\prime}\mathbf a(\mathbf k^\prime )$.
Using $
    \chi_{\mathbf{k}}v_{\mathbf{k}i}^{\mathrm{qp}}
    =
    -\,
    \partial_{k_i}n_F(\epsilon^{\mathrm{qp}}_{\mathbf{k}}-\mu)
$
and integrating by parts over the Brillouin zone  yields
\begin{equation}
    \sigma_{xy}
    =
    \sigma^{\mathrm{qp}}_{xy}
    =
    \frac{e^{2}}{\hbar}
    \intk
    n_F(\epsilon^{\mathrm{qp}}_{\mathbf{k}}-\mu)
    \Omega^{\mathrm{qp}}_{xy}(\mathbf{k}).
    \label{eq:AHE-final-compact}
\end{equation}
The AHE is therefore the Berry-curvature integral of the
quasiparticle band, with no additional Landau-backflow factor.

The transverse thermal Hall effect follows from introducing Luttinger's
gravitational sources \cite{Luttinger1964}. We introduce a gravitational vector potential
$\mathbf{A}^{g}$ (conjugate to the heat current) which encodes the gravitoelectric field
$\mathbf{E}^{g}=-\boldsymbol{\nabla}T/T$. This gives the anomalous potential
$-(\epsilon^{\mathrm{qp}}_{\mathbf{k}}-\mu)\,\mathbf{E}^{g}\cdot\mathbf{a}(\mathbf{k})$, where
the transported energy plays the role of the charge. Then, the
longitudinal heat current is locked to the charge current, 
$\bm{\mathcal{J}}^{Q}_{L}(\mathbf k)=\sum_{\mathbf k^\prime}\left[(1+f\chi)_\mathbf{\mathbf k \mathbf k^\prime}\,(\epsilon^{\mathrm{qp}}_\mathbf{k^\prime}-\mu)\,\mathbf{v}^{\mathrm{qp}}_\mathbf{k^\prime} \right]$.
The intrinsic thermal Hall conductivity $\kappa_{xy}$ is in turn obtained from the
transverse heat-current response to the gravitoelectric field, i.e., the
antisymmetric $\mathbf A^{g}$--$\mathbf E^{g}$ response obtained from the free energy~\cite{SM}
\begin{equation}
    \kappa_{xy}
    =
    \frac{k_{B}^{2}T}{\hbar}
    \intk
    \Omega_{xy}^{\rm qp}(\mathbf{k})\,
    I_{2}\!\left(
    \frac{\epsilon_{\mathbf{k}}^{\rm qp}-\mu}{T}
    \right),
    \label{eq:kappaxy}
\end{equation}
where $
    I_{2}(a)
    =
    \int_{a}^{\infty}dy\,
    y^{2}(-\frac{\partial n_{F}}{\partial y}).
$
Since $I_{2}(-\infty)=\pi^{2}/3$, comparison with
Eq.~\eqref{eq:AHE-final-compact} gives the transverse Wiedemann-Franz (WF) law 
$
    \frac{\kappa_{xy}}{T\sigma_{xy}}
    \longrightarrow
    L_{0}
    =
    \frac{\pi^{2}k_{B}^{2}}{3e^{2}}
$
as $T\rightarrow0$
\cite{XiaoYaoFangNiu2006,BergmanOganesyan2010}.

\emph{\color{blue} Drude weight.---} On an isotropic Fermi surface the
physical longitudinal current of Eq.~\eqref{eq:ward-main} reduces to
$\mathcal{J}_{L,i}(\mathbf{k})=ev^{\mathrm{qp}}_{\mathbf{k}i}(1+F_{1}/d)$,
with $F_{1}$ the $\ell=1$ Landau parameter.  With the quasiparticle current
$e\mathbf v^{\mathrm{qp}}$, we obtain the Landau form
$
    D_{xx}=\left(1+\frac{F_{1}}{d}\right)D^{\mathrm{qp}}_{xx}$,
where  $D^{\mathrm{qp}}_{xx}=\pi e^{2}\int_{\mathbf k}\chi_{\mathbf k}
    [v^{\mathrm{qp}}_{x}(\mathbf k)]^{2}$
is the Drude weight associated with the quasiparticle dispersion.
The total Drude weight can be decomposed into conventional and QG
contributions,
\begin{equation}
    D_{xx}
    =
    D^{\mathrm{con}}_{xx}
    +
    D^{\mathrm{QG}}_{xx},
    \label{eq:D-decomposition}
\end{equation}
with
$
    D_{xx}^{\mathrm{con}}
    =
    \frac{v^{0}_{F}}{v^{\mathrm{qp}}_{F}}
    D^{\mathrm{qp}}_{xx},
$
and
$
    D_{xx}^{\mathrm{QG}}
    =
    \frac{F^{\mathrm{QG}}_{1}}{d}
    D^{\mathrm{qp}}_{xx}.
$
Here $v_F^0$ and $v_F^{\mathrm{qp}}$ are the bare and quasiparticle Fermi velocities, respectively. We define the QG Landau parameter
$
    \frac{F^{\mathrm{QG}}_{1}}{d}
    \equiv
    \frac{1}{v^{\mathrm{qp}}_{F}}
    \frac{1}{V}
    \sum_{\mathbf{k}'}
    \hat{\mathbf{k}}\cdot
    \Lambda_{\mathbf{k}\mathbf{k}'}
    n^{0}_{\mathbf{k}'}
    |_{k_F},
$
which is the $\ell=1$ component on the Fermi surface. Charge conservation then fixes the total through the sum rule
$
    1+\frac{F_1}{d}
    =
    \frac{v_F^0}{v_F^{\mathrm{qp}}}
    +
    \frac{F_1^{\mathrm{QG}}}{d}
$.
{
In a conventional FL whose Landau function depends only on the momentum
transfer, the Ward identity fixes the physical current $\bm{\mathcal J}_L(\mathbf k)=e\mathbf v^0_{\mathbf k}$.
In a BL-FL, this relation can be broken: the
quantum-geometric current in Eq.~\eqref{eq:JQGomega} exists already at $T=0$, and 
 is further modified by thermal redistribution.
 In a flat band with trivial QG ($v_F^{0}=0$), interactions can generate a finite quasiparticle velocity, but the quasiparticle current $e\mathbf v^\mathrm{qp}_{\mathbf k}$ 
is exactly cancelled by Landau backflow, yielding $D^\mathrm{cov}=0$.
By contrast, a flat band with QG can exhibit transport of  entirely quantum-geometric origin.   
A further feature of the BL-FL theory is a possible  enhancement of transport with increasing temperature, as more states around the Fermi energy can contribute to the interaction-induced quasiparticle current,  $
\bm{\gamma}^{(\mathrm{int})\omega}(\mathbf{k})
    = \frac{1}{V} \sum_{\mathbf{k}'} \bm{\Lambda}_{\mathbf{k}\mathbf{k}'}
    n^0_{\mathbf{k}'}$,  which in turn enters the physical current $\bm{\mathcal J}_L$ in Eq.~\eqref{eq:ward-main}. We demonstrate this effect below for flat and narrow bands.
}

\begin{figure}[t!]
    \includegraphics[width=\columnwidth]{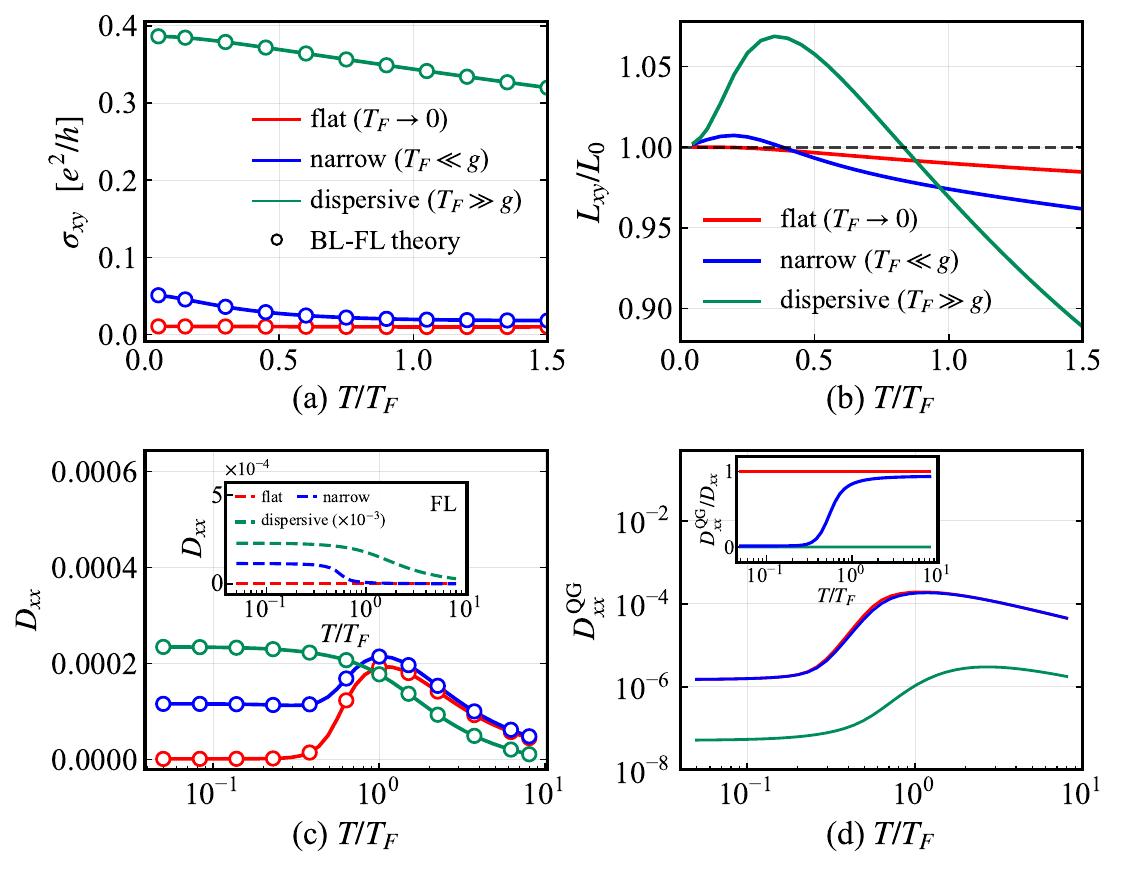}
    \caption{%
    Transport for three regimes in the Wilson-Dirac model: flat ($T_{F}\rightarrow 0$), narrow ($g/T_{F}=91$), and dispersive
    ($g/T_{F}=0.041$) from self-consistent finite-temperature HF.
    (a)~AHE: full self-consistent HF for the response from  twisting boundary conditions (solid) versus the Berry-curvature integral of the HF quasiparticle band (circles);  agreement to $3\times10^{-6}\,e^{2}/h$.
    (b)~Transverse Lorenz ratio $L_{xy}/L_{0}$ with the filled remote
    band restored; dashed line: $L_{0}$.
    (c)~Drude weight $D_{xx}$: Kohn stiffness from the twisted-boundary HF energy
    (solid) versus the prediction from BL-FL
    $D^{\rm qp}_{xx}[1+F_{1}(T)/2]$ (circles). Inset: conventional-FL controls; the flat-band control vanishes identically.
    (d)~QG contribution $D_{xx}^{\rm QG}$  in
    Eq.~\eqref{eq:D-decomposition}; inset: the fraction
    $D^{\rm QG}_{xx}/D_{xx}$. It equals one identically in the flat band,
    crosses over near $T\sim T_{F}$ in the narrow band, and
    is negligible in the dispersive band.
    Temperatures are normalized by the self-consistent Fermi scale
    $T_{F}$ of each branch.
    }
    \label{fig:fl-tfl-transport}
\end{figure}

\emph{\color{blue} Numerical results.---}
We test the BL-FL predictions using self-consistent finite-temperature
Hartree-Fock calculations for an isolated $|C|=1$ Wilson-Dirac band \cite{Wilson1974,QiWuZhang2006,Shen2012}
with screened density-density interactions of strength $g$~\cite{SM}.
We can independently
control the bare bandwidth and the separation from the remote band
while keeping  the quantum
geometry fixed. 
The isolated-band limit is  realized by 
$
    \Delta\gg g,\,T,\,W_{\mathrm{HF}},
$
where $W_{\mathrm{HF}}$ is the bandwidth of the self-consistent HF band.
Within HF, the quantum-geometric current in Eq.~\eqref{eq:JQGomega} is recognized by identifying the projected
exchange kernel $\Sigma^{\mathrm{HF}}_{\mathbf{k}\mathbf{k}'}$ with the
Landau function $f_{\mathbf{k}\mathbf{k}'}$. The static HF self-energy
defines the interacting Bloch basis,
$
h_{\mathrm{HF}}(\mathbf{k})
|u^{\mathrm{HF}}_{\mathbf{k}}\rangle
=
\epsilon^{\mathrm{HF}}_{\mathbf{k}}
|u^{\mathrm{HF}}_{\mathbf{k}}\rangle,
$
so that the band projection becomes
$
c_{\mathbf{k}\alpha}
\rightarrow
u^{\mathrm{HF}}_{\mathbf{k}\alpha}f_{\mathbf{k}}.
$
The projected exchange kernel then generates the interaction-induced current
$
\gamma_i^{\mathrm{(int),HF}}(\mathbf{k})
=
\frac{e}{V}
\sum_{\mathbf{k}'}
n^0_{\mathbf{k}'}
\left(
\nabla_{\mathbf{k}}
+
\nabla_{\mathbf{k}'}
\right)_i
\Sigma^{\mathrm{HF}}_{\mathbf{k}\mathbf{k}'},
$
which is the HF realization of the quantum-geometric current in
Eq.~\eqref{eq:JQGomega}.

We study three regimes: flat ($T_F\rightarrow0$), narrow ($g/T_F = 91$), and
dispersive ($g/T_F\simeq0.04$). Each is compared with a conventional-FL control
having the same dispersion, gap, filling, and interaction, but with the
Bloch-overlap factor $\langle u_{\mathbf{k}}|u_{\mathbf{k}'}\rangle$ in the
projected interaction kernel replaced by unity.

Fig.~\ref{fig:fl-tfl-transport}(a) confirms the AHE  in 
Eq.~\eqref{eq:AHE-final-compact} within BL-FL theory: the self-consistent HF boundary-twist
response, reconverged at each twist [see End Matter], agrees with the
Berry-curvature integral of the thermal quasiparticle band to
$3\times10^{-6}e^{2}/h$ for all regimes and temperatures. Panel~(b) confirms
the transverse WF law with the filled remote Chern band restored: all three
ratios approach $L_{0}$ at zero temperature, while their finite-temperature behavior depends strongly on the band dispersion.
The flat-band case remains closest to $L_{0}$ over a wide temperature range, decreasing only weakly to about $0.985L_{0}$ at $T=1.5T_{F}$.

Panels (c,d) test the predictions on the Drude weight. 
We obtain the Drude weight from the Kohn stiffness, i.e., the curvature of the free energy with respect to a boundary twist. 
It  agrees well with the relation from the BL-FL theory
 at all plotted temperatures.
For the flat band with trivial QG, the Drude weight remains identically zero in  conventional FL theory,
whereas the flat and narrow BL-FLs exhibit a finite thermal enhancement of the Drude weight, which peaks near $T\sim T_F$.  This enhancement originates from the quantum-geometric current in Eq.~\eqref{eq:JQGomega}.  
At low temperature, the occupied states form a small pocket around a band minimum, where both the quasiparticle velocity and the interaction-induced QG current are small. 
As temperature increases, thermal broadening allows more states to contribute to the QG current, enhancing the Drude weight.
When $T>T_F$, the occupation becomes nearly uniform 
$n_{\mathbf{k}}\rightarrow\nu$, and the Drude weight crosses over to  $D_{xx}\rightarrow\pi
    e^{2}\nu(1-\nu)\langle
    v^{\mathrm{qp}}_{x}\mathcal{J}_{L,x}/e\rangle_{\mathrm{BZ}}/k_{B}T$, which is consistent with the numerics.
Panel (d) isolates the QG contribution to the Drude weight: the fraction
$D_{xx}^{\mathrm{QG}}/D_{xx}$ is identically $1$ in the flat-band limit, crosses over
from $0.013$ to $0.94$ near $T\sim T_{F}$ in the narrow band, and stays below
$10^{-4}$ in the dispersive band.

\emph{\color{blue} Discussion.---}
We characterize two transport signatures in BL-FL theory: an intrinsic AHE determined by the quasiparticle Berry curvature and a QG contribution to the Drude weight.
These responses highlight different aspects of  quantum geometry: the AHE is controlled by the Berry curvature, whereas the longitudinal response reflects the interaction-induced QG current.
The effects are especially pronounced in narrow bands, where
interactions dominate over the band dispersion, allowing the QG current to become a leading contribution to transport. Narrow-band moir\'e systems therefore provide a natural setting in which these effects may be observable.

The framework also suggests several extensions. Spin or band degeneracy
promotes the quantum geometry to a non-Abelian structure and can modify spin-
and orbital-dependent Landau interactions and their associated instabilities.
The same construction can be applied to other probes, including strain and
rotation, and to collective modes such as zero sound. Disorder and interband
processes provide further corrections beyond the present isolated-band
treatment. More broadly, the BL-FL provides a low-energy framework in which
quasiparticle occupations, Landau interactions, and quantum geometry enter on
an equal footing. This becomes particularly important in narrow and flat bands,
where the conventional band velocity ceases to be the dominant scale for transport
and interaction-induced QG currents can take over.

\section*{Acknowledgments}

We thank B. Douçot, Z.-Y. Weng, T.-K. Ng, Y. Zhou, X.-Q. Sun and J.-Y. Chen for helpful
discussions. This work was in part supported by the Deutsche Forschungsgemeinschaft via the cluster of excellence ctd.qmat (EXC 2147, Project No.~390858490).

\bibliography{ref}

\begin{thebibliography}{88}
\bibitem [{gau()}]{gauge_note}%
  \BibitemOpen
  \href@noop {} {}\bibinfo {note} {Two gauge transformations act on the
  projection. First, $S_{\lambda}$ is invariant under the simultaneous rotation
  $|u_{\mathbf{k}}\rangle\to e^{i\theta(\mathbf{k})}|u_{\mathbf{k}}\rangle$,
  $f_{\mathbf{k}}\to e^{i\theta(\mathbf{k})}f_{\mathbf{k}}$: this redundancy
  ties the projected fermion to the Bloch frame, so a moving frame transfers
  its Berry phase to $f_{\mathbf{k}}$ and thereby generates the anomalous
  coupling. Second, under a smooth reparametrization $|u_{\mathbf{k}}\rangle\to
  e^{i\phi(\mathbf{k})}|u_{\mathbf{k}}\rangle$ at fixed $f_{\mathbf{k}}$, the
  anomalous coupling changes only by a total derivative along the quasiparticle
  trajectory, contributing a boundary term to the action and leaving the
  physical response unchanged---the action is gauge invariant.}\BibitemShut
  {Stop}%
\end{thebibliography}

\section*{End Matter}

\emph{\color{blue}Generic probes and QG currents.---}
A slow probe or weak perturbation introduces two quantum-geometric objects: an anomalous potential
$\dot{\mathbf k}\cdot\mathbf a(\mathbf k)$ and a corresponding interaction-induced QG current. Table~\ref{tab} summarizes the forces, anomalous couplings, current vertices, and representative responses.

To treat these probes in a single treatment, let $\lambda_{\mathbf k}$ denote the
generalized single-particle `charge' carried by a quasiparticle. In the presence
of the conjugate vector potential $\mathbf A^\lambda$, the momentum carried by
each fermionic state is shifted according to $ \mathbf k\rightarrow \mathbf
    k+\lambda_{\mathbf k}\mathbf A^\lambda . $ For charge transport
$\lambda_{\mathbf k}=e$; for heat transport $\lambda_{\mathbf
        k}=\epsilon^{\mathrm{qp}}_{\mathbf k}-\mu$; and for a momentum current
$\lambda_{\mathbf k}=k_j$. Expanding the projected interaction to linear order
in $\mathbf A^\lambda$ gives the interaction-induced vertex
\begin{equation}
    \mathcal{J}^{\lambda,\mathrm{int}}_{\alpha}(\mathbf{k})
    =
    \frac{1}{V}\sum_{\mathbf{k}'}
    \left[
        \left(
        \lambda_{\mathbf{k}}\partial_{k_{\alpha}}
        +
        \lambda_{\mathbf{k}'}\partial_{k'_{\alpha}}
        \right)
        f_{\mathbf{k}\mathbf{k}'}
        \right]
    n_{\mathbf{k}'},
    \label{eq:EM-vertex}
\end{equation}
which separates naturally into center-of-mass and relative-momentum derivatives,
\begin{equation}
    \!\!\!\!\!\!\!\mathcal{J}^{\lambda}_{\alpha}(\mathbf{k})
    \!=\!\!
    \frac{1}{V}\sum_{\mathbf{k}'} n_{\mathbf{k}'}
    \!\left(
    \frac{\lambda_{\mathbf{k}}+\lambda_{\mathbf{k}'}}{2}
    \mathfrak D_{\alpha}
    +
    \frac{\lambda_{\mathbf{k}}-\lambda_{\mathbf{k}'}}{2}
    \mathfrak d_{\alpha}
    \right)\!
    f_{\mathbf{k}\mathbf{k}'}
    ,
    \label{eq:EM-master}
\end{equation}
where
$\mathfrak D_{\alpha}
    =\partial_{k_{\alpha}}+\partial_{k'_{\alpha}}$
and
$\mathfrak d_{\alpha}
    =\partial_{k_{\alpha}}-\partial_{k'_{\alpha}}$
act only on the interaction kernel. The $\mathfrak D_{\alpha}$ term probes the center-of-mass momentum dependence of the projected interaction. By contrast, the $\mathfrak d_{\alpha}$ term describes the conventional relative-momentum, or backflow, channel. For charge transport, $\lambda_{\mathbf k}=e$ is momentum independent, so the
relative-momentum term vanishes identically and Eq.~\eqref{eq:EM-master}
reduces to $\bm{\gamma}^{(\mathrm{int}),\omega}$ of Eq.~\eqref{eq:JQGomega}.
The derivations and descriptions for the probes listed in
Table~\ref{tab} are given in the Supplemental Material~\cite{SM}.

\emph{\color{blue}Numerical method.---}
All data in Fig.~\ref{fig:fl-tfl-transport} are obtained from
self-consistent finite-temperature HF calculations on a $121\times 121$
momentum grid ($221\times 221$ for the flat-band case). Temperature is
swept from high to low, with each converged solution used to initialize
the next, until the HF residual falls below $2\times10^{-12}$.
We define the Fermi scale of each branch as
$T_F=\mu-\min_{\mathbf k}\epsilon^{\mathrm{qp}}_{\mathbf k}$.
The full AHE in Fig.~\ref{fig:fl-tfl-transport}(a) is evaluated from the
many-body Berry curvature of the self-consistent HF state under
boundary twists. At each temperature, we first determine the equilibrium occupations $n_\mathbf{k}^0$ from the HF approximation for the system with periodic boundary conditions. The occupations are then kept fixed while HF mean-field parameters are self-consistently reconverged under the boundary twists $(\theta_x,\theta_y)=(\pm\delta\theta/2,\pm\delta\theta/2)$, with
$\delta\theta=0.1$. Denoting the resulting upper-band projectors by
$P_{1,\ldots,4}(\mathbf k)$, the gauge-invariant Berry flux associated with
each momentum label is $ \phi_{\mathbf k}^{\mathrm{full}} =
    \arg\operatorname{Tr} \left[ P_1(\mathbf k)P_2(\mathbf k) P_3(\mathbf
        k)P_4(\mathbf k) \right]. $ The Hall conductivity is then
\begin{equation}
    \sigma_{xy}^{\mathrm{full}}
    =
    \frac{e^2}{h}\,
    \frac{2\pi}{\delta\theta^2}
    \sum_{\mathbf k}
    n_{\mathbf k}^{(0)}
    \phi_{\mathbf k}^{\mathrm{full}} .
    \label{eq:sigma-full-twist}
\end{equation}
In turn, $\sigma_{xy}^{\mathrm{qp}}$ associated with the quasiparticle band is obtained from this formula 
with the projectors under twisted boundary conditions constructed using the HF mean-field parameters determined under periodic boundary conditions.

The Kohn stiffness in Fig.~\ref{fig:fl-tfl-transport}(c) is extracted from the
curvature of the free energy with respect to a boundary twist $ D_{xx} = \pi e^2 \left.
    \partial_{A_x}^{2}F_{\mathrm{HF}}(A_x) \right|_{A_x\to0}, $
where $A_x$ is the uniform  vector potential corresponding to the twist in the $x$ direction.
The curvature is evaluated using a fourth-order five-point stencil.
The result agrees with the independent linear-response
calculation to relative accuracy
$7\times10^{-5}$ for all branches and temperatures.

\emph{\color{blue}  Drude weight at finite temperature.---}
The temperature dependence of the Drude weight can be understood from its
energy-resolved transport weight,
\begin{equation}
    D_{xx}(T)
    =
    \pi e
    \int d\varepsilon
    \mathcal N_{\Phi}(\varepsilon)
    \left(
    -\frac{\partial n_F}{\partial\varepsilon}
    \right),
    \label{eq:D-spectral}
\end{equation}
where
$
    \mathcal N_{\Phi}(\varepsilon)
    =
    \int_{\mathbf{k}}
    \delta(
    \varepsilon-\epsilon^{\mathrm{qp}}_{\mathbf{k}}
    )
        v^{\mathrm{qp}}_{x}(\mathbf{k})
        \mathcal J_{L,x}(\mathbf{k}).
$
For the Wilson-Dirac model in the flat-band limit, $\mathcal N_{\Phi}\propto k^{2}$ near the band minimum, where the QG current is weak. 
Thermal broadening can yield three regimes as the temperature increases.
At low temperature, we obtain the Sommerfeld behavior 
$
    D(T)\simeq
    D(0)\left[1+a(T/T_F)^2\right]
$ with $
    D(0)\propto k_F^2
$.
For $ T\agt  T_F$, the contribution from states within the energy window $\Delta \varepsilon$ around the Fermi energy  gives
$
    D(T)
    \simeq
    \frac{A}{k_BT}
    e^{-\Delta\varepsilon/k_BT}
$, where $\Delta \varepsilon\sim k_B T_F$ is the characteristic excitation energy of the states dominating transport.
This implies a maximum at $k_B T^*\sim \Delta_B$ in $D(T)$. At even higher temperature, the occupations become nearly uniform and the Drude weight crosses over to the $1/T$ decay as discussed in the main text.

\begin{table*}[!b]
    \caption{
    Generic forces, anomalous potential, and
    interaction-induced  QG currents in a BL-FL theory. A perturbation that drives the quasiparticle
    momentum according to $\dot{\mathbf k}$ generates the anomalous potential $\dot{\mathbf k}\cdot {\mathbf a}(\mathbf k)$ where ${\mathbf a}(\mathbf k)$ is, respectively, the Berry connection of the quasiparticle (bare particle) band in an interacting (non-interacting) system. 
    For a probe coupling to a quasiparticle with generalized charge $\lambda_\mathbf{k}$, the interaction-induced QG current is given in Eq.~\eqref{eq:EM-master} in the BL-FL theory.
    In the table, as detailed in the Supplemental Material~\cite{SM}, $\varphi(\mathbf r)$ is a smooth single-particle potential energy;
    $u_{ij}$ denotes strain with elastic gauge potentials
    $\mathbf A_{\rm el}$ and $\Phi_{\rm el}$;
    $\omega_z$ is the local frame-rotation rate; and 
    $s_z^{\mathbf k}=\langle u_\mathbf k|S_z^{\rm orbital}|u_\mathbf{k}\rangle$ is the projection of the orbital rotation generator onto the quasiparticle band.
    Furthermore, 
    $\Gamma_{ij}=\dot e_i^{a}e_{ja}$ whose symmetric and antisymmetric parts describe strain and
    frame rotation, respectively; and $T^{a}_{0i}$ denotes torsion generated
    by a time-dependent vielbein gradient.
    }
    \label{tab} {\scriptsize{}%
        \renewcommand{\arraystretch}{1.5}%
        \arrayrulecolor[gray]{0.6}%
        \setlength{\tabcolsep}{7pt}%
        \begin{tabular}{lllll}
            \toprule
            {\scriptsize\tcell{1.2in}{}}                                                        &
            {\scriptsize\tcell{1.35in}{Generalized force $\dot{\bk}$}}                          &
            {\scriptsize\tcell{1.5in}{ Anomalous potential }}                                   &
            {\scriptsize\tcell{1.45in}{ Charge $\lambda_{\bk}$  in QG current}}                     &
            {\scriptsize\tcell{1.25in}{Representative response}}\tabularnewline
            \midrule
            {\scriptsize\tcell{0.85in}{Electric field}}                                         &
            {\scriptsize\tcell{1.35in}{$-e\bE$}}                                                &
            {\scriptsize\tcell{1.5in}{$-e\bE\cdot\ba(\bk)$}}                                    &
            {\scriptsize\tcell{1.45in}{$e$ }}                          &
            {\scriptsize\tcell{1.25in}{AHE \cite{Nagaosa2010,Haldane2004}                         \\
            electric polarization \cite{KingSmithVanderbilt1993}                                  \\
                        Drude weight}}\tabularnewline
            \cmidrule(lr){1-5}
            {\scriptsize\tcell{0.85in}{Magnetic field}}                                         &
            {\scriptsize\tcell{1.35in}{$-e\dot{\br}\times\bB$,
            $\dot{\br}=\nabla_{\bk}\epsilon^{\mathrm{qp}}_{\bk}$}}                              &
            {\scriptsize\tcell{1.5in}{$-e(\dot{\br}\times\bB)\cdot\ba(\bk)$}}                   &
            {\scriptsize\tcell{1.45in}{$e$}}                                                 &
            {\scriptsize\tcell{1.25in}{magnetization \cite{Thonhauser2005,XiaoShiNiu2005},
                        Streda response \cite{Streda1982}, cyclotron
                        dynamics}}\tabularnewline
            \cmidrule(lr){1-5}
            {\scriptsize\tcell{0.85in}{Scalar potential}}                                       &
            {\scriptsize\tcell{1.35in}{$-\nabla_{\br}\varphi(\br)$}}                            &
            {\scriptsize\tcell{1.5in}{$-\nabla\varphi\cdot\ba(\bk)$}}                           &
            {\scriptsize\tcell{1.45in}{$e$
            }}                                                                                  &
            {\scriptsize\tcell{1.25in}{AHE
                    }}\tabularnewline
            \cmidrule(lr){1-5}
            {\scriptsize\tcell{1.1in}{Thermal}}                                                 &
            {\scriptsize\tcell{1.35in}{$(\epsilon^{\mathrm{qp}}_{\bk}-\mu)\frac{\nabla T}{T}$}} &
            {\scriptsize\tcell{1.5in}{$(\epsilon^{\mathrm{qp}}_{\bk}-\mu) \frac{\nabla T}{T}
            \cdot\ba(\bk)$}}                                                                    &
            {\scriptsize\tcell{1.45in}{$\epsilon^{\mathrm{qp}}_{\bk}-\mu$
            }}                                                                                  &
            {\scriptsize\tcell{1.25in}{thermal Hall
                        \cite{Luttinger1964,QinNiuShi2011,KarmakarEtAl2022,MangeolleSavaryBalents2024}, thermoelectricity}}\tabularnewline
            \cmidrule(lr){1-5}
            {\scriptsize\tcell{0.85in}{Strain                                                     \\
            acoustic field}}                                                                    &
            {\scriptsize\tcell{1.35in}{$-\partial_{t}\bA_{\rm el}
            -\nabla\Phi_{\rm el}$, or locally $-\dot{u}_{ij}k_{j}$}}                            &
            {\scriptsize\tcell{1.5in}{$(-\partial_{t}\bA_{\rm el}
            -\nabla\Phi_{\rm el})\cdot\ba(\mathbf k)$}}                                                    &
            {\scriptsize\tcell{1.45in}{$k_{j}$}}                                                &
            {\scriptsize\tcell{1.25in}{piezoelectric
                        \cite{KingSmithVanderbilt1993,2000JPCS...61..147V}, elastoconductivity \cite{2025PhRvB.111r4408T},
                        phonon drag \cite{Vozmediano2010,GuineaKatsnelsonGeim2010}} }\tabularnewline
            \cmidrule(lr){1-5}
            {\scriptsize\tcell{0.85in}{Frame rotation spin connection}}                         &
            {\scriptsize\tcell{1.35in}{$-\omega_{z}\hat{\mathbf{z}}\times\bk$
            with $D_{t}=\partial_{t}-i\omega_{z}S^{\rm basis}_{z}$}}                            &
            {\scriptsize\tcell{1.5in}{$-\omega_{z}(\hat{\mathbf{z}}\times\bk)
            \cdot\ba +\omega_{z}s_z(\mathbf k)$}}                                             &
            {\scriptsize\tcell{1.45in}{$(\hat{\mathbf{z}}\times\bk)_{j}$}}                      &
            {\scriptsize\tcell{1.25in}{orbital spin,                                              \\
                        Hall viscosity  \cite{WenZee1992,AvronSeilerZograf1995,Read2009}, rotational susceptibility}}\tabularnewline
            \cmidrule(lr){1-5}
            {\scriptsize\tcell{0.85in}{Metric or shear quench}}                                 &
            {\scriptsize\tcell{1.35in}{$-\Gamma_{ij}k_{j}$}}                                    &
            {\scriptsize\tcell{1.5in}{$-\Gamma_{ij}k_{j}a_{i}(\bk)$}}                           &
            {\scriptsize\tcell{1.45in}{$k_{j}$}}                                                &
            {\scriptsize\tcell{1.25in}{nematic susceptibility
                        \cite{OganesyanKivelsonFradkin2001,FradkinEtAl2010},
                        dynamical Hall viscosity in
                        optical response
                        \cite{HoyosSon2012,BradlynGoldsteinRead2012}}}\tabularnewline
            \cmidrule(lr){1-5}
            {\scriptsize\tcell{0.85in}{Torsion/vielbein gradient}}                              &
            {\scriptsize\tcell{1.35in}{$-T^{a}_{\ 0i}k_{a}$ in the local
            frame}}                                                                             &
            {\scriptsize\tcell{1.5in}{$-T^{a}_{\ 0i}k_{a}a_{i}(\bk)$}}                          &
            {\scriptsize\tcell{1.45in}{$(\epsilon^{\mathrm{qp}}_{\bk}-\mu)\,k_{a}$}}            &
            {\scriptsize\tcell{1.25in}{torsional viscosity
                        \cite{HughesLeighFradkin2011}, energy magnetization
                        \cite{QinNiuShi2011}}}\tabularnewline
            \bottomrule
        \end{tabular}}
    \renewcommand{\arraystretch}{1}%
\end{table*}

\clearpage
\onecolumngrid
\setcounter{equation}{0}
\setcounter{figure}{0}
\setcounter{table}{0}
\setcounter{section}{0}
\renewcommand{\theequation}{S\arabic{equation}}
\renewcommand{\thefigure}{S\arabic{figure}}
\renewcommand{\thetable}{S\arabic{table}}
\renewcommand{\thesection}{S\arabic{section}}

\begin{center}
{\large\textbf{Supplemental Material for}}\\[4pt]
{\large\textbf{``Berry-Landau Fermi-liquid theory: transport in presence of quantum geometry''}}\\[6pt]
Shuai A.~Chen and Roderich Moessner\\[2pt]
{\small Max Planck Institute for the Physics of Complex Systems, N\"othnitzer Stra\ss e 38, 01187 Dresden, Germany}
\end{center}

\makeatletter\let\addcontentsline\arxiv@addcontentsline\makeatother
\tableofcontents


\section{Generalized Peierls substitution}

\subsection{Grassmann-valued Lagrange-multiplier projection}

For completeness and consistency, we revisit the generalized Peierls
substitution approach introduced in Ref.~\cite{xkxw-1134} for a
non-interacting system. We consider a system with an active band separated
from the remote bands by a large energy gap. The band structure is
described by $N_\mathrm{orb}$-orbital Hamiltonian $H_{0}=\sum_{\mathbf{k}}h(\mathbf{k})$ with 
\begin{equation}
h(\mathbf{k})=\sum_{\alpha\beta}c^{\dagger}_{\mathbf{k}\alpha}t_{\alpha\beta}(\mathbf{k})c_{\mathbf{k}\beta}.
\end{equation}
Here 
$c^\dagger _{\mathbf{k}\alpha}$ creates a fermion on orbital $\alpha$ at momentum $\mathbf k$.
In particular,
we denote the Bloch wave at the crystal momentum $\mathbf{k}$ as
$|u_{\mathbf{k}}\rangle$: $h(\mathbf{k})|u_{\mathbf{k}}\rangle=\epsilon_{\mathbf{k}}|u_{\mathbf{k}}\rangle$.
We can project the whole system onto the Hilbert subspace $\mathbb{H}^{\mathrm{sub}}$
of the active band. According to Ref.~\cite{xkxw-1134}, the band
projection can be realized by means of the Grassmann-valued Lagrange
multipliers $\lambda_{\mathbf{k}\alpha}$ and $\lambda^{\dagger}_{\mathbf{k}\alpha}$
\begin{equation}
S_{\lambda}=\int^{\beta}_{0}d\tau\sum_{\mathbf{k}\alpha}\left[\lambda^{\dagger}_{\mathbf{k}\alpha}(\tau)(c_{\mathbf{k}\alpha}(\tau)-u_{\mathbf{k}\alpha}(\tau)f_{\mathbf{k}}(\tau))+(c^{\dagger}_{\mathbf{k}\alpha}(\tau)-u^{*}_{\mathbf{k}\alpha}(\tau)f^{\dagger}_{\mathbf{k}}(\tau))\lambda_{\mathbf{k}\alpha}(\tau)\right],\label{eq:Slambda-sm}
\end{equation}
where the operator $f_{\mathbf{k}}$ annihilates a fermion at the
momentum $\mathbf{k}$ with the Bloch wave $|u_{\mathbf{k}}\rangle$.
The partition function in imaginary time can be formulated as 
\begin{equation}
Z_{0}=\int Dc^{\dagger}DcDf^{\dagger}DfD\lambda^{\dagger}D\lambda e^{-S_{0}-S_{\lambda}},\label{eq:zfree}
\end{equation}
where $S_{0}=\int d\tau\left[\sum_{\mathbf{k}\alpha}c^{\dagger}_{\mathbf{k}\alpha}(\tau)\partial_{\tau}c_{\mathbf{k}\alpha}(\tau)+H_{0}\right]$.
We can obtain the band projection $c_{\mathbf{k}\alpha}=u_{\mathbf{k}\alpha}f_{\mathbf{k}}$
by integrating out the Lagrange multipliers $\lambda_{\mathbf{k}\alpha}$
and $\lambda^{\dagger}_{\mathbf{k}\alpha}$. Within the Hilbert subspace
$\mathbb{H}^{\mathrm{sub}}$, all fields $\lambda$, $f$ and $c$
carry $U(1)$ charge. The band projector is modified in the presence
of an electromagnetic field. For instance, the Bloch wave becomes
time dependent as its crystal momentum follows the semiclassical dynamical
equation (see Sec. \ref{subsec:Houston-ansatz-for}).

\subsection{Band projection under interactions }

\label{subsec:variational-u}

We show here how the Bloch wave entering the band-projection constraint
is determined in the interacting theory. We first formulate the Baym-Kadanoff
functional for the full single-particle Green function and then
restrict it to the low-energy quasiparticle subspace through the constraint
$S_{\lambda}$.

For convenience, we use $i=(\tau_{i},\mathbf{k}_{i},\alpha_{i})$
to collect the imaginary time, momentum, and orbital indices. The
Euclidean action is given by $S[c^{\dagger},c]=-c^{\dagger}_{1}(G^{-1}_{0})_{12}c_{2}+S_{\mathrm{int}}[c^{\dagger},c]$,
where repeated indices are summed or integrated over. Here $G_{0}$
is the bare Green function with 
\begin{equation}
G^{-1}_{0}(i\omega_{n},\mathbf{k})=i\omega_{n}+\mu-h_{0}(\mathbf{k}).
\end{equation}
We introduce the source term $J_{12}$ to the partition function,
\begin{equation}
Z[J]=\int D[c^{\dagger},c]\exp\left[-S[c^{\dagger},c]-c^{\dagger}_{1}J_{12}c_{2}\right].
\end{equation}
Formally, we can obtain the Green function $G_{12}=\frac{\delta W}{\delta J_{12}}$,
with $W[J]=-\ln Z[J]$. The Baym-Kadanoff functional $\Gamma[G]$
is obtained from the Legendre transformation 
\begin{equation}
\Gamma[G]=W[J]-\mathrm{Tr}(JG),
\end{equation}
which satisfies $\frac{\delta\Gamma}{\delta G}=-J$. The full single-particle
Green function corresponds to the stationary point $\frac{\delta\Gamma[G]}{\delta G}=0$.
For the non-interacting system, we have $G^{-1}=G^{-1}_{0}-J$. From
\begin{align}
Z_{0}[J] & =\int D[c^{\dagger},c]\exp\left[-c^{\dagger}\left(G^{-1}_{0}-J\right)c\right]\propto\det\left[G^{-1}_{0}-J\right],
\end{align}
we obtain 
\begin{equation}
\Gamma_{0}[G]=-\ln Z_{0}[J]-\mathrm{Tr}(JG)=\mathrm{Tr}\ln G-\mathrm{Tr}\left[\left(G^{-1}_{0}-G^{-1}\right)G\right].
\end{equation}
Interactions add the Luttinger-Ward functional $\Phi[G]$, giving
\begin{equation}
\Gamma[G]=\mathrm{Tr}\ln G-\mathrm{Tr}\left[\left(G^{-1}_{0}-G^{-1}\right)G\right]+\Phi[G].
\end{equation}
Here $\Phi[G]$ is the sum of closed two-particle-irreducible diagrams
constructed from the full Green function $G$ and bare interaction
vertices. Equivalently, $\Phi[G]$ may be defined non-perturbatively
as the interaction part of the two-particle-irreducible effective
action. The self-energy can be deduced via $\Sigma_{12}=\frac{\delta\Phi[G]}{\delta G_{12}}$.

We impose the constraint $S_{\lambda}$ on the Baym-Kadanoff functional
$\Gamma[G]$ by integrating over the Lagrange multipliers $\lambda^{\dagger}$
and $\lambda$. Writing $k=(i\omega_{n},\mathbf{k})$, we decompose
the full Green function 
\begin{equation}
G(k)=G_{P}(k)+G_{Q}(k),\label{eq:G=00003D00003DP+Q}
\end{equation}
with $G_{P}=PGP$ and $G_{Q}=G-G_{P}$, where $P$ denotes the band
projector onto the active band. The intraband component $G_{P}$ possesses
the structure $G_{P}(k)=g(k)P_{\mathbf{k}}$ with $g(k)=\langle f_{k}f^{\dagger}_{k}\rangle$
and $P_{\mathbf{k}}=|u_{\mathbf{k}}\rangle\langle u_{\mathbf{k}}|$.
Physically, $G_{P}$ ($G_{Q}$) is the coherent (incoherent) part
of the Green function. In the isolated-band limit, $G_{Q}$ vanishes
and thus $G(k)=G_{P}(k)$.

We restrict the Baym-Kadanoff functional to the Hilbert subspace,
\begin{equation}
\tilde{\Gamma}[g,P]\equiv\Gamma(G=gP).
\end{equation}
The constraint $S_{\lambda}$ does not provide an additional contribution
to $\Phi[G]$. After the band projection, the Luttinger-Ward functional
$\Phi[G]$ becomes $\Phi_{\mathrm{proj}}[g,P]=\Phi[G=gP]$. We are now
ready to determine the parameters $g(k)$ and $P_{\mathbf{k}}$. First,
we vary the scalar propagator $g(k)$ while fixing $P_{\mathbf{k}}$.
From $\delta G(k)=P_{\mathbf{k}}\delta g(k)$ and $G^{-1}=g^{-1}P+G^{-1}_{Q}$,
we have the variation 
\begin{equation}
\delta\Gamma=\sum_{k}\delta g(k)\mathrm{Tr}\left[P(G^{-1}-(G^{-1}_{0}-\Sigma))\right]=0.\label{eq:deltaG/deltaG}
\end{equation}
Given the isolated-band condition, Eq. (\ref{eq:deltaG/deltaG}) reduces
to 
\begin{equation}
\frac{\delta\tilde{\Gamma}}{\delta g(k)}=g^{-1}(k)-\langle u_{\mathbf{k}}|(G^{-1}_{0}-\Sigma)|u_{\mathbf{k}}\rangle=0.
\end{equation}
Thus, we obtain the scalar quasiparticle propagator $g^{-1}(k)=\langle u_{\mathbf{k}}|G^{-1}|u_{\mathbf{k}}\rangle$.
Second, we consider the variation of the Bloch wave $u_{\mathbf{k}}$.
The normalization $\langle u_{\mathbf{k}}|u_{\mathbf{k}}\rangle=1$
indicates that the physically relevant variation is perpendicular
to $u_{\mathbf{k}}$. Thus, we can represent the variation as an infinitesimal
rotation $P_{\mathbf{k}}\rightarrow e^{\eta_{\mathbf{k}}}P_{\mathbf{k}}e^{-\eta_{\mathbf{k}}}$,
which gives $\delta P_{\mathbf{k}}=[\eta_{\mathbf{k}},P_{\mathbf{k}}]$.
Equivalently, we can write $|\delta u_{\mathbf{k}}\rangle=Q_{\mathbf{k}}\eta_{\mathbf{k}}|u_{\mathbf{k}}\rangle$
up to an irrelevant change of the phase of $u_{\mathbf{k}}$. We apply
the same infinitesimal transformation to Eq. (\ref{eq:G=00003D00003DP+Q}),
$G\rightarrow e^{\eta}Ge^{-\eta}$, or equivalently, $\delta_{u}G=[\eta,G]$.
Therefore, we have 
\begin{align}
\delta_{u}\Gamma & =\mathrm{Tr}\left[(G^{-1}-(G^{-1}_{0}-\Sigma))[\eta,G]\right]\nonumber \\
 & =-\mathrm{Tr}[\eta[G,G^{-1}_{0}-\Sigma]]
\end{align}
where we used the identity $\mathrm{Tr}[G^{-1}[\eta,G]]=\mathrm{Tr}[G^{-1}\eta G-G^{-1}G\eta]=0$.
Stationarity with respect to arbitrary orbital rotations $\eta$ gives
$[G,G^{-1}_{0}-\Sigma]=0$. From $G(k)|u_{\mathbf{k}}\rangle=g(k)P_{\mathbf{k}}|u_{\mathbf{k}}\rangle=g(k)|u_{\mathbf{k}}\rangle$,
we obtain the eigenvalue problem, 
\begin{equation}
G(G^{-1}_{0}-\Sigma)|u_{\mathbf{k}}\rangle=(G^{-1}_{0}-\Sigma)G|u_{\mathbf{k}}\rangle=g(k)(G^{-1}_{0}-\Sigma)|u_{\mathbf{k}}\rangle.
\end{equation}
This implies that $(G^{-1}_{0}-\Sigma)|u_{\mathbf{k}}\rangle$ is
also an eigenvector of $G$ with eigenvalue $g(k)$. Due to the non-degeneracy
of the quasiparticle eigenvalues, we further simplify the eigenvalue
problem 
\begin{equation}
(G^{-1}_{0}-\Sigma)|u_{\mathbf{k}}\rangle=g^{-1}(k)|u_{\mathbf{k}}\rangle.
\end{equation}
This gives the condition for Bloch waves in the presence of interactions.
Let $z_{\mathbf{k}}$ denote the quasiparticle pole, $g^{-1}(z_{\mathbf{k}},\mathbf{k})=0$.
We arrive at the condition used in the main text, $G^{-1}(z_{\mathbf{k}},\mathbf{k})|u_{\mathbf{k}}\rangle=0$,
with $G^{-1}=G^{-1}_{0}-\Sigma$. For a long-lived quasiparticle,
$z_{\mathbf{k}}=\xi^{\mathrm{qp}}_{\mathbf{k}}-i\Gamma_{\mathbf{k}}$
with $\Gamma_{\mathbf{k}}\rightarrow0$, the anti-Hermitian part of
the self-energy $\Sigma$ can be neglected to leading order.

The role of the remote bands can be shown explicitly. Writing the
inverse Green function in the block form, 
\begin{equation}
G^{-1}=\begin{pmatrix}K_{PP} & K_{PQ}\\
K_{QP} & K_{QQ}
\end{pmatrix},\qquad|u^{{\rm qp}}\rangle=\begin{pmatrix}u_{P}\\
u_{Q}
\end{pmatrix},
\end{equation}
the equation $G^{-1}(z_{\mathbf{k}},\mathbf{k})|u_{\mathbf{k}}\rangle=0$
leads to $K_{QP}u_{P}+K_{QQ}u_{Q}=0,$ and thus $u_{Q}=-K^{-1}_{QQ}K_{QP}u_{P}.$
With the gap $\Delta$ between the active and remote bands, we have $K^{-1}_{QQ}\sim\frac{1}{\Delta}$
and $K_{QP}\sim U$ with $U$ the interaction scale, so that $u_{Q}\sim\frac{U}{\Delta}u_{P}.$
Thus, interactions allow hybridization between the active and remote
bands; the hybridization is small only when $U/\Delta\ll1$. Eliminating
$u_{Q}$ gives the exact active-sector pole equation $\left[K_{PP}-K_{PQ}K^{-1}_{QQ}K_{QP}\right]u_{P}=0$.

\subsection{Houston Ansatz for the Bloch wave on an accelerated trajectory }

\label{subsec:Houston-ansatz-for}

We consider how the band projector in Eq.~(\ref{eq:Slambda-sm}) changes
in response to a weak static electric field. We choose the velocity
gauge $\mathbf{A}(t)=-\mathbf{E}t$ for the static electric field
(here $t$ is real time). Then, we have the Hamiltonian $H_{0}(\mathbf{k},t)=H_{0}(\mathbf{k}+e\mathbf{A}(t))$.
The Houston Ansatz tells us that the solution to the instantaneous
Schrödinger equation $H_{0}(\mathbf{k}+e\mathbf{A}(t))|u_{n\mathbf{k}}(t)\rangle=\epsilon_{n}(\mathbf{k},t)|u_{n\mathbf{k}}(t)\rangle$
(with $n$ the band index) takes the form $|u_{n\mathbf{k}}(t)\rangle=|u_{n\mathbf{k}(t)}\rangle$,
where the crystal momentum $\mathbf{k}$ follows the classical dynamics
$\dot{\mathbf{k}}(t)=-e\mathbf{E}$.

For convenience, we restate the Houston Ansatz in imaginary time.
After the Wick rotation $t=-i\tau$, the velocity gauge becomes $\mathbf{A}(\tau)=-(-i\tau)\mathbf{E}=i\mathbf{E}\tau$.
Therefore, the imaginary-time Hamiltonian in the velocity gauge is
$H_{E}(\tau)=H_{0}(\mathbf{k}+ie\mathbf{E}\tau)$. Accordingly, we
have the imaginary-time Houston Ansatz $|u_{n\mathbf{k}}(\tau)\rangle=|u_{n\mathbf{k}(\tau)}\rangle$
with the crystal momentum following the evolution equation $d\mathbf{k}/d\tau=-e\mathbf{E}$.

\subsection{Gravitomagnetic Peierls substitution}

To study thermal transport, we couple the system to a background gravitomagnetic
vector potential $\bm{\lambda}$.

\subsubsection{Energy twist from the gravitomagnetic field}

We place the system on a Euclidean torus with the metric $ds^{2}=(d\tau+i\lambda_{i}dx^{i})^{2}+\delta_{ij}dx^{i}dx^{j}$.
Here the off-diagonal metric component $g_{0i}=i\lambda_{i}$ is a gravitomagnetic
vector potential, conjugate to the energy current. The gravitomagnetic
background appears through the global identifications of the Euclidean
torus, 
\begin{equation}
(\tau,\mathbf{x})\sim(\tau+\beta,\mathbf{x})\sim(\tau+i\lambda_{i}L_{i},\mathbf{x}+L_{i}\mathbf{e}_{i}),
\end{equation}
 with $\mathbf{e}_{i}$ the unit vector along the $x^{i}$-direction.
The introduction of the gravitomagnetic field twists the boundary
condition. This is the gravitational analogue of a flux-induced boundary
twist.

In imaginary time, the Bloch wave is written as $|\psi_{\mathbf{k}}\rangle=e^{-\epsilon(\mathbf{k})\tau}e^{i\mathbf{k}\cdot\mathbf{x}}|u_{\mathbf{k}}\rangle$.
Under the gravitomagnetic twisted boundary condition $(\tau,\mathbf{x})\sim(\tau+i\lambda_{i}L_{i},\mathbf{x}+L_{i}\mathbf{e}_{i})$,
the state $|\psi_{\mathbf{k}}\rangle$ acquires a factor $e^{-i\lambda_{i}L_{i}\epsilon(\mathbf{k})}e^{iL_{i}k_{i}}$
after winding once around the torus. Single-valuedness gives the quantization
condition for the momentum $\mathbf{k}$ 
\begin{align}
(k_{i}-\epsilon(\mathbf{k})\lambda_{i})L_{i} & =2\pi n_{i}.
\end{align}
We define the twisted momentum $\mathbf{k}^{\lambda}=\mathbf{k}+\epsilon(\mathbf{k}^{\lambda})\bm{\lambda}$;
the corresponding energy is $\epsilon^{\lambda}(\mathbf{k})=\epsilon(\mathbf{k}^{\lambda})=\epsilon(\mathbf{k}+\bm{\lambda}\epsilon(\mathbf{k}^{\lambda}))$.
This is the gravitomagnetic analogue of the usual Peierls substitution.
To linear order in $\boldsymbol{\lambda}$, $\mathbf{k}^{\lambda}=\mathbf{k}+\bm{\lambda}\epsilon(\mathbf{k})+\mathcal{O}(\lambda^{2})$,
such that $\mathbf{k}\rightarrow\mathbf{k}+\epsilon(\mathbf{k})\bm{\lambda}$.
We will derive the quantum-geometric thermal current through the introduction
of the gravitomagnetic vector potential.

\subsubsection{Quantum-geometric thermal current}

To derive the quantum-geometric thermal current, we solve the dynamics
of the crystal momentum. We consider the Lagrangian for the gravitomagnetic
coupling with $\boldsymbol{\lambda}=\boldsymbol{\lambda}(t)$: $L=(\mathbf{k}+\epsilon^{\lambda}(\mathbf{k})\boldsymbol{\lambda})\cdot\dot{\mathbf{r}}-\epsilon^{\lambda}(\mathbf{k})$
with the canonical momentum $\bm{\pi}=\frac{\partial L}{\partial\dot{\mathbf{r}}}=\mathbf{k}+\epsilon^{\lambda}(\mathbf{k})\boldsymbol{\lambda}$.
From the Euler-Lagrange equation $\frac{d}{dt}\left(\mathbf{k}+\epsilon^{\lambda}(\mathbf{k})\boldsymbol{\lambda}\right)=0$,
the crystal momentum $\mathbf{k}$ follows the equation of motion
$d_{t}\mathbf{k}=-\epsilon(\mathbf{k})d_{t}\boldsymbol{\lambda}$
to first order in $\boldsymbol{\lambda}$. This gives the overlap
\begin{align}
-i\langle u_{\mathbf{k}}|\partial_{\tau}u_{\mathbf{k}}\rangle & =d_{t}\mathbf{k}\cdot\mathbf{a}(\mathbf{k})\nonumber \\
 & =-\epsilon_{\mathbf{k}}d_{t}\boldsymbol{\lambda}\cdot\mathbf{a}(\mathbf{k})\equiv\epsilon_{\mathbf{k}}\mathbf{E}^{g}\cdot\mathbf{a}(\mathbf{k}),
\end{align}
where the gravitoelectric field $\mathbf{E}^{g}=-d_{t}\boldsymbol{\lambda}$
plays the role of the temperature gradient. We
next determine the first-order perturbation to the Bloch wave due
to the gravitomagnetic field. To linear order in $\boldsymbol{\lambda}$,
the gravitomagnetic Peierls substitution gives $H^{\lambda}(\mathbf{k})=H(\mathbf{k})+\sum_{\alpha}\lambda_{\alpha}j^{E}_{\alpha}(\mathbf{k})+\mathcal{O}(\lambda^{2})$
with the energy-current operator $j^{E}_{\alpha}=\frac{1}{2}\{\partial_{\alpha}H(\mathbf{k}),H(\mathbf{k})\}$.
Expanding the Bloch wave as $|u^{\lambda}_{\mathbf{k}}\rangle=|u_{\mathbf{k}}\rangle+\lambda_{\alpha}|\delta^{\lambda}_{\alpha}u_{\mathbf{k}}\rangle+\mathcal{O}(\lambda^{2})$,
perturbation theory gives 
\begin{align}
|\delta^{\lambda}_{\alpha}u_{\mathbf{k}}\rangle & =\sum_{m}|u_{m\mathbf{k}}\rangle\frac{\langle u_{m\mathbf{k}}|j^{E}_{\alpha}(\mathbf{k})|u_{\mathbf{k}}\rangle}{\epsilon_{\mathbf{k}}-\epsilon_{m\mathbf{k}}}=\frac{\epsilon_{\mathbf{k}}+H(\mathbf{k})}{2}(1-|u_{\mathbf{k}}\rangle\langle u_{\mathbf{k}}|)|\partial_{\alpha}u_{\mathbf{k}}\rangle,
\end{align}
with $\langle u_{m\mathbf{k}}|j^{E}_{\alpha}(\mathbf{k})|u_{\mathbf{k}}\rangle=\frac{1}{2}(\epsilon^{2}_{\mathbf{k}}-\epsilon^{2}_{m\mathbf{k}})\langle u_{m\mathbf{k}}|\partial_{\alpha}u_{\mathbf{k}}\rangle.$
We can then construct the corresponding terms in the energy functional
of the BL-FL theory. Explicitly, we have 
\begin{equation}
\delta E[n,\lambda]=\sum_{\mathbf{k}}\xi^{\mathrm{qp}}_{\mathbf{k}}\mathbf{E}^{g}\cdot\mathbf{a}(\mathbf{k})+\frac{1}{2V}\sum_{\alpha}\lambda_{\alpha}\left(\mathcal{J}^{Q,\mathrm{QG}}_{\alpha}+\mathcal{J}^{Q,\mathrm{backflow}}_{\alpha}\right),
\end{equation}
with 
\begin{align}
\mathcal{J}^{Q,\mathrm{QG}}_{\alpha} & =\frac{1}{4V}\sum_{\mathbf{kk}^{\prime}}(\xi_{\mathbf{k}}+\xi_{\mathbf{k}^{\prime}})(\partial_{k_{\alpha}}+\partial_{k^{\prime}_{\alpha}})f_{\mathbf{k}\mathbf{k}^{\prime}}n_{\mathbf{k}}n_{\mathbf{k}^{\prime}},\\
\mathcal{J}^{Q,\mathrm{backflow}}_{\alpha} & =\frac{1}{4V}\sum_{\mathbf{kk}^{\prime}}(\xi_{\mathbf{k}}-\xi_{\mathbf{k}^{\prime}})(\partial_{k_{\alpha}}-\partial_{k^{\prime}_{\alpha}})f_{\mathbf{k}\mathbf{k}^{\prime}}n_{\mathbf{k}}n_{\mathbf{k}^{\prime}}.
\end{align}

\subsection{Quantum-geometric current and Hartree-Fock approximation}

We apply the Hartree-Fock approximation to the interactions and to the quantum-geometric
current operator. Starting from the density-density interaction 
\begin{equation}
H_{\mathrm{int}}=\frac{1}{2V}\sum_{\mathbf{p}\mathbf{p}^{\prime}\mathbf{q}\alpha\beta}V_{\mathbf{q}}c^{\dagger}_{\mathbf{p}+\mathbf{q}\alpha}c^{\dagger}_{\mathbf{p}^{\prime}-\mathbf{q}\beta}c_{\mathbf{p}^{\prime}\beta}c_{\mathbf{p}\alpha},
\end{equation}
the Hartree-Fock approximation gives an HF Hamiltonian $H^{\mathrm{HF}}_{0}(\mathbf{k})=\sum_{\alpha\beta}t^{\mathrm{HF}}_{\alpha\beta}(\mathbf{k})c^{\dagger}_{\mathbf{k}\alpha}c_{\mathbf{k}\beta}$
with 
\[
t^{\mathrm{HF}}_{\alpha\beta}(\mathbf{k})=t_{\alpha\beta}(\mathbf{k})-\frac{1}{V}\sum_{\mathbf{k}^{\prime}}V_{\mathbf{k}-\mathbf{k}^{\prime}}\langle c^{\dagger}_{\mathbf{k}^{\prime}\alpha}c_{\mathbf{k}^{\prime}\beta}\rangle.
\]
$H^{\mathrm{HF}}_{0}(\mathbf{k})$ defines the Bloch states of
the quasiparticles in the active HF band: 
\begin{align}
H^{\mathrm{HF}}_{0}(\mathbf{k})|u^{\mathrm{HF}}_{\mathbf{k}}\rangle & =\epsilon^{\mathrm{HF}}_{\mathbf{k}}|u^{\mathrm{HF}}_{\mathbf{k}}\rangle,\\
\mathbf{a}^{\mathrm{HF}}(\mathbf{k}) & =-i\langle u^{\mathrm{HF}}_{\mathbf{k}}|\nabla_{\mathbf{k}}|u^{\mathrm{HF}}_{\mathbf{k}}\rangle.
\end{align}
After projection onto the active HF band, the interaction term becomes,
\begin{equation}
H_{\mathrm{int}}=\frac{1}{2V}\sum_{\mathbf{p}\mathbf{p}^{\prime}\mathbf{q}}V_{\mathbf{q}}\langle u^{\mathrm{HF}}_{\mathbf{p}+\mathbf{q}}|u^{\mathrm{HF}}_{\mathbf{p}}\rangle\langle u^{\mathrm{HF}}_{\mathbf{p}^{\prime}-\mathbf{q}}|u^{\mathrm{HF}}_{\mathbf{p}^{\prime}}\rangle f^{\dagger}_{\mathbf{p}+\mathbf{q}}f^{\dagger}_{\mathbf{p}^{\prime}-\mathbf{q}}f_{\mathbf{p}^{\prime}}f_{\mathbf{p}}.\label{eq:proj_Hint}
\end{equation}

We introduce a weak static electric field through the gauge potential
$\mathbf{A}(\tau)=i\mathbf{E}\tau$. Within the Houston Ansatz, the
Hartree-Fock Bloch wave follows $|u^{\mathrm{HF}}_{\mathbf{p}}(\tau)\rangle=|u^{\mathrm{HF}}_{\mathbf{p}(\tau)}\rangle$
with $\mathbf{\dot{p}}(\tau)=-e\mathbf{E}$. After integrating out
the Lagrange multipliers $\lambda,\lambda^{\dagger}$ and the fermion
fields $c,c^{\dagger}$, the interaction term becomes coupled to the
gauge potential; to leading order in $\mathbf{A}$, 
\begin{align}
H_{\mathrm{int}}[A]= & H_{\mathrm{int}}+\frac{1}{2V}\sum_{\mathbf{p}\mathbf{p}^{\prime}\mathbf{q}}\int_{\omega}eA_{\alpha}(\omega)e^{-i\omega t}V_{\mathbf{q}}\mathfrak{D}_{\alpha}\left[\langle u^{\mathrm{HF}}_{\mathbf{p}+\mathbf{q}}|u^{\mathrm{HF}}_{\mathbf{p}}\rangle\langle u^{\mathrm{HF}}_{\mathbf{p}^{\prime}-\mathbf{q}}|u^{\mathrm{HF}}_{\mathbf{p}^{\prime}}\rangle\right]f^{\dagger}_{\mathbf{p}+\mathbf{q}}f^{\dagger}_{\mathbf{p}^{\prime}-\mathbf{q}}f_{\mathbf{p}^{\prime}}f_{\mathbf{p}}+\mathcal{O}(A^{2}).\label{eq:Hint=00003D00005BA=00003D00005D}
\end{align}
Here we define a derivative operator $\mathfrak{D}_{\alpha}$, satisfying
\begin{align}
\mathfrak{D}_{\alpha}(\langle u_{\mathbf{p}+\mathbf{q}}|u_{\mathbf{p}}\rangle) & =\langle\mathcal{D}_{\alpha}u_{\mathbf{p}+\mathbf{q}}|u_{\mathbf{p}}\rangle+\langle u_{\mathbf{p}+\mathbf{q}}|\mathcal{D}_{\alpha}u_{\mathbf{p}}\rangle,\\
\mathfrak{D}_{\alpha}\left[\langle u_{\mathbf{p}+\mathbf{q}}|u_{\mathbf{p}}\rangle\langle u_{\mathbf{p}^{\prime}-\mathbf{q}}|u_{\mathbf{p}^{\prime}}\rangle\right] & =\mathfrak{D}_{\alpha}(\langle u_{\mathbf{p}+\mathbf{q}}|u_{\mathbf{p}}\rangle)\langle u_{\mathbf{p}^{\prime}-\mathbf{q}}|u_{\mathbf{p}^{\prime}}\rangle+\langle u_{\mathbf{p}+\mathbf{q}}|u_{\mathbf{p}}\rangle\mathfrak{D}_{\alpha}(\langle u_{\mathbf{p}^{\prime}-\mathbf{q}}|u_{\mathbf{p}^{\prime}}\rangle),
\end{align}
with $\mathcal{D}_{\alpha}$ the covariant derivative $|\mathcal{D}_{\alpha}u_{\mathbf{p}}\rangle=(\partial_{p_{\alpha}}-ia_{\alpha}(\mathbf{p}))|u_{\mathbf{p}}\rangle$.
From Eq. (\ref{eq:Hint=00003D00005BA=00003D00005D}), we can identify
the quantum-geometric current operator 
\begin{equation}
J^{\mathrm{QG}}_{\alpha}(\mathbf{p})=\frac{1}{V}\sum_{\mathbf{p}^{\prime}\mathbf{q}}V_{\mathbf{q}}\mathfrak{D}_{\alpha}\left[\langle u^{\mathrm{HF}}_{\mathbf{p}+\mathbf{q}}|u^{\mathrm{HF}}_{\mathbf{p}}\rangle\langle u^{\mathrm{HF}}_{\mathbf{p}^{\prime}-\mathbf{q}}|u^{\mathrm{HF}}_{\mathbf{p}^{\prime}}\rangle\right]f^{\dagger}_{\mathbf{p}+\mathbf{q}}f^{\dagger}_{\mathbf{p}^{\prime}-\mathbf{q}}f_{\mathbf{p}^{\prime}}f_{\mathbf{p}}.
\end{equation}
The operator $J^{\mathrm{QG}}_{\alpha}(\mathbf{p})$ can be simplified
within the HF approximation. Explicitly, we have 
\begin{align}
 & \frac{1}{2}V_{\mathbf{q}}\mathfrak{D}_{\alpha}\left[\langle u^{\mathrm{HF}}_{\mathbf{p}+\mathbf{q}}|u^{\mathrm{HF}}_{\mathbf{p}}\rangle\langle u^{\mathrm{HF}}_{\mathbf{p}^{\prime}-\mathbf{q}}|u^{\mathrm{HF}}_{\mathbf{p}^{\prime}}\rangle\right]f^{\dagger}_{\mathbf{p}+\mathbf{q}}f^{\dagger}_{\mathbf{p}^{\prime}-\mathbf{q}}f_{\mathbf{p}^{\prime}}f_{\mathbf{p}}\nonumber \\
\rightarrow & \frac{1}{2}V_{\mathbf{0}}\mathfrak{D}_{\alpha}\left[\langle u^{\mathrm{HF}}_{\mathbf{p}}|u^{\mathrm{HF}}_{\mathbf{p}}\rangle\langle u^{\mathrm{HF}}_{\mathbf{p}^{\prime}}|u^{\mathrm{HF}}_{\mathbf{p}^{\prime}}\rangle\right]\left[n^{\mathrm{HF}}_{\mathbf{p}}f^{\dagger}_{\mathbf{p}^{\prime}}f_{\mathbf{p}^{\prime}}+n^{\mathrm{HF}}_{\mathbf{p}^{\prime}}f^{\dagger}_{\mathbf{p}}f_{\mathbf{p}}\right]\nonumber \\
 & -\frac{1}{2}V_{\mathbf{p}-\mathbf{p}^{\prime}}\mathfrak{D}_{\alpha}\left[\langle u^{\mathrm{HF}}_{\mathbf{p}^{\prime}}|u^{\mathrm{HF}}_{\mathbf{p}}\rangle\langle u^{\mathrm{HF}}_{\mathbf{p}}|u^{\mathrm{HF}}_{\mathbf{p}^{\prime}}\rangle\right]\left[n^{\mathrm{HF}}_{\mathbf{p}^{\prime}}f^{\dagger}_{\mathbf{p}}f_{\mathbf{p}}+n^{\mathrm{HF}}_{\mathbf{p}}f^{\dagger}_{\mathbf{p}^{\prime}}f_{\mathbf{p}^{\prime}}\right]\nonumber \\
 & +\frac{1}{2}V_{\mathbf{q}}\mathfrak{D}_{\alpha}\left[\langle u^{\mathrm{HF}}_{\mathbf{p}+\mathbf{q}}|u^{\mathrm{HF}}_{\mathbf{p}}\rangle\langle u^{\mathrm{HF}}_{\mathbf{p}^{\prime}-\mathbf{q}}|u^{\mathrm{HF}}_{\mathbf{p}^{\prime}}\rangle\right]:f^{\dagger}_{\mathbf{p}+\mathbf{q}}f^{\dagger}_{\mathbf{p}^{\prime}-\mathbf{q}}f_{\mathbf{p}^{\prime}}f_{\mathbf{p}}:\nonumber \\
= & -V_{\mathbf{p}-\mathbf{p}^{\prime}}\mathfrak{D}_{\alpha}\left[\langle u^{\mathrm{HF}}_{\mathbf{p}^{\prime}}|u^{\mathrm{HF}}_{\mathbf{p}}\rangle\langle u^{\mathrm{HF}}_{\mathbf{p}}|u^{\mathrm{HF}}_{\mathbf{p}^{\prime}}\rangle\right]n^{\mathrm{HF}}_{\mathbf{p}^{\prime}}f^{\dagger}_{\mathbf{p}}f_{\mathbf{p}}\nonumber \\
 & +\frac{1}{2}V_{\mathbf{q}}\mathfrak{D}_{\alpha}\left[\langle u^{\mathrm{HF}}_{\mathbf{p}+\mathbf{q}}|u^{\mathrm{HF}}_{\mathbf{p}}\rangle\langle u^{\mathrm{HF}}_{\mathbf{p}^{\prime}-\mathbf{q}}|u^{\mathrm{HF}}_{\mathbf{p}^{\prime}}\rangle\right]:f^{\dagger}_{\mathbf{p}+\mathbf{q}}f^{\dagger}_{\mathbf{p}^{\prime}-\mathbf{q}}f_{\mathbf{p}^{\prime}}f_{\mathbf{p}}:,
\end{align}
where the normal-ordered term $:f^{\dagger}f^{\dagger}ff:$ denotes
the residual interactions beyond the HF approximation. Thus, $H_{\mathrm{int}}[A]$
in Eq. (\ref{eq:Hint=00003D00005BA=00003D00005D}) simplifies to 
\begin{align}
H_{\mathrm{int}}[A] & =H_{\mathrm{int}}-\frac{e}{V}\sum_{\mathbf{p}^{\prime}}V_{\mathbf{p}-\mathbf{p}^{\prime}}\mathfrak{D}_{\alpha}\left[|\langle u^{\mathrm{HF}}_{\mathbf{p}}|u^{\mathrm{HF}}_{\mathbf{p}^{\prime}}\rangle|^{2}\right]n^{\mathrm{HF}}_{\mathbf{p}^{\prime}}A_{\alpha}(\omega)f^{\dagger}_{\mathbf{p}}(p_{0}+\omega)f_{\mathbf{p}}(p_{0}).
\end{align}
For the coefficients 
\begin{align}
V_{\mathbf{p-p^{\prime}}}\mathfrak{D}_{\alpha}\left[\left|\langle u^{\mathrm{HF}}_{\mathbf{p}}|u^{\mathrm{HF}}_{\mathbf{p}^{\prime}}\rangle\right|^{2}\right]= & (\partial_{p_{\alpha}}+\partial_{p^{\prime}_{\alpha}})\left[V_{\mathbf{p-p^{\prime}}}\left|\langle u^{\mathrm{HF}}_{\mathbf{p}}|u^{\mathrm{HF}}_{\mathbf{p}^{\prime}}\rangle\right|^{2}\right]=-(\partial_{p_{\alpha}}+\partial_{p^{\prime}_{\alpha}})\Sigma^{\mathrm{HF}}_{\mathbf{p}\mathbf{p}^{\prime}},
\end{align}
with the self-energy $\Sigma^{\mathrm{HF}}_{\mathbf{p}\mathbf{p}^{\prime}}=-V_{\mathbf{p-p^{\prime}}}\left|\langle u^{\mathrm{HF}}_{\mathbf{p}}|u^{\mathrm{HF}}_{\mathbf{p}^{\prime}}\rangle\right|^{2}$,
we obtain a compact expression for the quantum-geometric current operator
\begin{align}
J^{\mathrm{QG}}_{\alpha}(\mathbf{p}) & =eT\sum_{p_{0}}\int_{\mathbf{p}^{\prime}}n^{\mathrm{HF}}_{\mathbf{p}^{\prime}}(\partial_{p_{\alpha}}+\partial_{p^{\prime}_{\alpha}})\Sigma^{\mathrm{HF}}_{\mathbf{p}\mathbf{p}^{\prime}}f^{\dagger}_{\mathbf{p}}(p_{0}+\omega)f_{\mathbf{p}}(p_{0}).\label{eq:JQG-op-def}
\end{align}

\section{Intrinsic AHE in a non-interacting system }

In this section, we provide details of the derivation of the intrinsic
AHE in the non-interacting system. Within the Houston Ansatz $|u_{\mathbf{k}}(\tau)\rangle\equiv|u_{\mathbf{k}(\tau)}\rangle$,
the crystal momentum follows the semiclassical equation of motion
$\dot{\mathbf{k}}(\tau)=e\dot{\mathbf{A}}(\tau)=-e\mathbf{E}(\tau)$.
After imposing the band projection, we have the Lagrangian density
$\mathcal{L}$ 
\begin{align}
\mathcal{L} & =f^{\dagger}(\mathbf{x},\tau)\left[\partial_{\tau}+\epsilon(-i\nabla+e\mathbf{A})-ie\mathbf{E}\cdot\mathbf{a}(-i\nabla)\right]f(\mathbf{x},\tau),\label{seq:L}
\end{align}
where formally $\mathbf{a}(-i\nabla)$ is understood as a pseudo-differential
operator by replacing $\mathbf{k}\rightarrow-i\nabla$. We regularize
a uniform static electric field $\mathbf{E}$ by a finite frequency
$\omega$ with $\mathbf{E}(\omega)=i\omega\mathbf{A}(\omega)$ and
take the static limit $\omega\rightarrow0$ at the end. We obtain
the effective action by integrating out the fermion fields, giving
\begin{align}
Z[A] & =\exp(-S_{\mathrm{eff}}[A]),\\
S_{\mathrm{eff}}[A] & =-\mathrm{Tr}\ln(G^{-1}_{0}-V[A]),
\end{align}
with 
\begin{align}
G^{-1}_{\mathbf{k}}(i\omega) & =i\omega-\epsilon(\mathbf{k}),\\
V[A] & =e\mathbf{A}\cdot\mathbf{v}(\mathbf{k})+e\mathbf{E}\cdot\mathbf{a}(\mathbf{k})=e\mathbf{A}\cdot\mathbf{v}(\mathbf{k})+ie\omega_{n}\mathbf{A}(i\omega_{n})\cdot\mathbf{a}(\mathbf{k}).
\end{align}
The polarization tensor is defined at quadratic order in $\mathbf{A}$
in the expansion of the effective action $S_{\mathrm{eff}}[A]$,
\begin{equation}
\Pi^{ij}(i\omega)=e^{2}T\sum_{\nu}\int\frac{d^{d}\mathbf{k}}{(2\pi)^{d}}\mathrm{Tr}\left[G_{\mathbf{k}}(i\nu)\gamma_{i}(k)G_{\mathbf{k}}(i\nu+i\omega)\gamma_{j}(k)\right].
\end{equation}
Here $\gamma_{i}(k)=\gamma^{(\mathrm{v})}_{i}(k)+\gamma^{(\mathrm{B})}_{i}(k)$
with $\gamma^{(\mathrm{v})}_{i}(k)=v_{i}(\mathbf{k})$ and $\gamma^{(\mathrm{B})}_{i}(k)=i\omega_{n}a_{i}(\mathbf{k})$.
The conductivity follows from the antisymmetric part of the polarization tensor,
$\sigma_{ij}=\lim_{\omega\rightarrow0}\Pi^{ij,a}(\omega+i\eta,\mathbf{0})/(i\omega)$
with $\Pi^{ij,a}=\frac{1}{2}(\Pi^{ij}-\Pi^{ji})$. For the Hall conductivity,
we calculate 
\begin{align}
\Pi^{(\mathrm{hall})}_{ij}(i\omega_{n},\mathbf{0}) & =e^{2}T\sum_{\nu}\int\frac{d^{d}\mathbf{k}}{(2\pi)^{d}}\mathrm{Tr}\left[G_{\mathbf{k}}(i\nu)\gamma^{(\mathrm{v})}_{i}G_{\mathbf{k}}(i\nu+i\omega_{n})\gamma^{(\mathrm{B})}_{j}\right]-(i\leftrightarrow j)\nonumber \\
 & =i\omega_{n}e^{2}T\sum_{\nu}\int\frac{d^{d}\mathbf{k}}{(2\pi)^{d}}\left[v_{i}(\mathbf{k})a_{j}(\mathbf{k})-v_{j}(\mathbf{k})a_{i}(\mathbf{k})\right]G_{\mathbf{k}}(i\nu)G_{\mathbf{k}}(i\nu+i\omega_{n})\nonumber \\
 & =i\omega_{n}e^{2}\int\frac{d^{d}\mathbf{k}}{(2\pi)^{d}}\left[v_{i}(\mathbf{k})a_{j}(\mathbf{k})-v_{j}(\mathbf{k})a_{i}(\mathbf{k})\right]\left[T\sum_{\nu}G^{2}_{\mathbf{k}}(i\nu)\right]+\mathcal{O}(\omega^{2}_{n})\nonumber \\
 & =-i\omega_{n}e^{2}\int\frac{d^{d}\mathbf{k}}{(2\pi)^{d}}\left[v_{i}(\mathbf{k})a_{j}(\mathbf{k})-v_{j}(\mathbf{k})a_{i}(\mathbf{k})\right]\frac{\partial n_{F}(\epsilon_{\mathbf{k}})}{\partial\epsilon_{\mathbf{k}}}+\mathcal{O}(\omega^{2}_{n}),
\end{align}
where we expand $G_{\mathbf{k}}(i\nu+i\omega_{n})$ at small $\omega_{n}$:
$G_{\mathbf{k}}(i\nu+i\omega_{n})=G_{\mathbf{k}}(i\nu)-i\omega_{n}G^{2}_{\mathbf{k}}(i\nu)+\mathcal{O}(\omega^{2}_{n})$
and use $T\sum_{\nu}G^{2}_{\mathbf{k}}(i\nu)=-\frac{\partial n_{F}(\epsilon_{\mathbf{k}})}{\partial\epsilon_{\mathbf{k}}}$.
Noting that 
\begin{align}
 & \int\frac{d^{d}\mathbf{k}}{(2\pi)^{d}}\left[v_{i}(\mathbf{k})a_{j}(\mathbf{k})-v_{j}(\mathbf{k})a_{i}(\mathbf{k})\right]\frac{\partial n_{F}(\epsilon_{\mathbf{k}})}{\partial\epsilon_{\mathbf{k}}}\nonumber \\
= & \int\frac{d^{d}\mathbf{k}}{(2\pi)^{d}}\left[\frac{\partial\epsilon_{\mathbf{k}}}{\partial k_{i}}a_{j}(\mathbf{k})-\frac{\partial\epsilon_{\mathbf{k}}}{\partial k_{j}}a_{i}(\mathbf{k})\right]\frac{\partial n_{F}(\epsilon_{\mathbf{k}})}{\partial\epsilon_{\mathbf{k}}}\label{seq:AHE_FS}\\
= & -\int\frac{d^{d}\mathbf{k}}{(2\pi)^{d}}n_{F}(\epsilon_{\mathbf{k}})\Omega_{ij}(\mathbf{k}),
\end{align}
we obtain the Hall conductivity 
\begin{align}
\sigma_{xy} & =\lim_{\omega\rightarrow0}\frac{1}{i\omega}\Pi^{xy,a}(\omega)=\frac{e^{2}}{\hbar}\int\frac{d^{d}\mathbf{k}}{(2\pi)^{d}}n_{F}(\epsilon_{\mathbf{k}})\Omega_{xy}(\mathbf{k}).
\end{align}

\section{Nozières-Luttinger construction}

In this section, we apply the Nozières-Luttinger construction to the
BL-FL theory.

We first review the construction, which
identifies the low-energy processes responsible for the
collective response of a Fermi liquid. Particle-hole pairs near the Fermi surface carrying
a small momentum $\mathbf{q}$ and frequency $\omega$ give rise to the singular dependence on
the frequency $\omega$ and momentum $\mathbf{q}$. In contrast, the
regular interaction processes are collected into an effective scattering
vertex $I$. Diagrammatically, $I$ is irreducible in the relevant
particle-hole channel: it cannot be split into two pieces by cutting
the propagators of the particle-hole pair. The full vertex $\Gamma$
satisfies the Bethe-Salpeter equation, 
\begin{equation}
\Gamma(\mathbf{k},\mathbf{k}^{\prime};i\omega,\mathbf{q})=I_{\mathbf{k}\mathbf{k}'}+\int_{\mathbf{p}}I_{\mathbf{k}\mathbf{p}}R(\mathbf{p};i\omega,\mathbf{q})\Gamma(\mathbf{p},\mathbf{k}^{\prime};i\omega,\mathbf{q}),
\end{equation}
where $R$ is the bare particle-hole kernel 
\begin{equation}
R(\mathbf{p};i\omega,\mathbf{q})=T\sum_{\nu}G(\mathbf{p}+\mathbf{q},i\nu+i\omega)G(\mathbf{p},i\nu)=\frac{n_{\mathbf{p}}-n_{\mathbf{p+\mathbf{q}}}}{i\omega+\epsilon_{\mathbf{p}}-\epsilon_{\mathbf{p}+\mathbf{q}}},\label{eq:ph-kernel}
\end{equation}
with $n_{\mathbf{p}}=n_{F}(\epsilon_{\mathbf{p}})$. For instance,
given the density-density interaction in Eq. (\ref{eq:proj_Hint}),
the bare particle-hole irreducible vertex is $I^{0}_{\mathbf{k}\mathbf{k}'}=-V_{\mathbf{k-k^{\prime}}}|\langle u_{\mathbf{k}}|u_{\mathbf{k}^{\prime}}\rangle|^{2}$,
while $I$ is obtained by including the interaction renormalization
of the bare vertex $I^{0}$.

In operator notation, the Bethe-Salpeter equation is $\Gamma=I+IR\Gamma,$
or equivalently $\Gamma=(1-IR)^{-1}I.$ Near $\left(\omega,\mathbf{q}\right)=(0,\mathbf{0})$,
the vertex $\Gamma$ depends on the order in which two limits are
taken: the dynamic limit 
\[
\Gamma^{\omega}(\mathbf{k},\mathbf{k}')=\lim_{\omega\rightarrow0}\lim_{\mathbf{q}\rightarrow0}\Gamma(\mathbf{k},\mathbf{k}';i\omega,\mathbf{q}),
\]
 and the static limit 
\[
\Gamma^{q}(\mathbf{k},\mathbf{k}')=\lim_{\mathbf{q}\rightarrow0}\lim_{\omega\rightarrow0}\Gamma(\mathbf{k},\mathbf{k}';i\omega,\mathbf{q}).
\]
 For the kernel defined in Eq. (\ref{eq:ph-kernel}), the dynamic
limit is $R^{\omega}(\mathbf{p})=\lim_{\omega\rightarrow0}\lim_{\mathbf{q}\rightarrow0}R(\mathbf{p};i\omega,\mathbf{q})=0.$
Thus, we have $\Gamma^{\omega}_{\mathbf{k}\mathbf{k}^{\prime}}=I_{\mathbf{k}\mathbf{k}^{\prime}}.$

The Landau interaction is defined from the vertex $\Gamma^{\omega}_{\mathbf{k}\mathbf{k}^{\prime}}$
in the dynamic limit 
\begin{equation}
f_{\mathbf{k}\mathbf{k}'}=Z_{\mathbf{k}}Z_{\mathbf{k}'}\Gamma^{\omega}_{\mathbf{k}\mathbf{k}^{\prime}},\label{eq:Landau-from-Gammaomega}
\end{equation}
where $Z_{\mathbf{k}}$ is the quasiparticle weight. In the static
limit, the particle-hole kernel becomes 
\begin{align*}
R^{q}(\mathbf{p}) & =\lim_{\mathbf{q}\rightarrow0}\lim_{\omega\rightarrow0}R(\mathbf{p};i\omega,\mathbf{q})=\partial_{\epsilon}n_{F}(\epsilon_{\mathbf{p}})\equiv-\chi_{\mathbf{p}}.
\end{align*}
The Bethe-Salpeter equation becomes $\Gamma^{q}=I-I\chi\Gamma^{q}$.
Introducing the renormalized static vertex $\tilde{\Gamma}^{q}_{\mathbf{kk}^{\prime}}=Z_{\mathbf{k}}Z_{\mathbf{k}'}\Gamma^{q}_{\mathbf{k}\mathbf{k}^{\prime}}$,
and using Eq. (\ref{eq:Landau-from-Gammaomega}), we obtain 
\[
\tilde{\Gamma}^{q}_{\mathbf{kk}^{\prime}}=f_{\mathbf{k}\mathbf{k}'}-\int_{\mathbf{p}}f_{\mathbf{k}\mathbf{p}}\chi_{\mathbf{p}}\tilde{\Gamma}^{q}_{\mathbf{p}\mathbf{k}^{\prime}}.
\]
 Thus $f_{\mathbf{k}\mathbf{k}'}$ is the interaction between quasiparticles
in the dynamic limit, while the static vertex contains additional
dressing from quasiparticles near the Fermi surface.

\subsection{Bethe-Salpeter equations for the electromagnetic vertex}

A subtlety arises because the interaction generates not only the particle-hole
ladder but also an interaction-induced contribution to the current operator. 
Thus, the quantum-geometric current forms part of the irreducible
electromagnetic vertex.

Starting from the full Green function in an external vector potential,
$G^{-1}[A]=G^{-1}_{0}[A]-\Sigma[G,A]$, the electromagnetic vertex
is obtained by differentiating with respect to $\mathbf{A}$, 
\begin{equation}
\frac{\delta G^{-1}[A]}{\delta A_{i}}=\frac{\delta G^{-1}_{0}[A]}{\delta A_{i}}-\left(\frac{\delta\Sigma[G,A]}{\delta A_{i}}\right)_{G}-\left(\frac{\delta\Sigma[G,A]}{\delta G}\right)_{A}\frac{\delta G}{\delta A_{i}}.
\end{equation}
The anomalous potential term will be discussed later. Defining the
full electromagnetic vertex $\Lambda_{i}=-\frac{\delta G^{-1}[A]}{\delta A_{i}}$,
and using $\frac{\delta G}{\delta A_{i}}=G\Lambda_{i}G$, we have
the Bethe-Salpeter equation 
\[
\Lambda_{i}=\gamma^{(\mathrm{v})}_{i}+\gamma^{(\mathrm{int})}_{i}+IR\Lambda_{i}.
\]
Here $\gamma^{(\mathrm{v})}_{i}(\mathbf{k})=-\frac{\delta G^{-1}_{0}[A]}{\delta A_{i}}=e\partial_{k_{i}}\epsilon^{0}_{\mathbf{k}}$
is the bare velocity vertex, $\gamma^{(\mathrm{int})}_{i}=\left(\frac{\delta\Sigma[G,A]}{\delta A_{i}}\right)_{G}$
is the interaction-induced irreducible quantum-geometric current vertex,
$I=\left(\frac{\delta\Sigma[G,A]}{\delta G}\right)_{A}$ is the particle-hole
irreducible four-point vertex, and $R$ is the bare particle-hole
kernel.

We first solve the Bethe-Salpeter equation in the dynamic limit. Since
$R^{\omega}=0$, we have $\Lambda_{i}=\gamma^{(\mathrm{v})}_{i}+\gamma^{(\mathrm{int})}_{i}$.
With $f_{\mathbf{k}\mathbf{k}'}=Z_{\mathbf{k}}Z_{\mathbf{k}'}\Gamma^{\omega}_{\mathbf{k}\mathbf{k}^{\prime}}$,
the interaction-induced quantum-geometric current vertex becomes $\gamma^{(\mathrm{int},\omega)}_{i}=e\int_{\mathbf{k}^{\prime}}n_{\mathbf{k}^{\prime}}(\partial_{k_{i}}+\partial_{k^{\prime}_{i}})f_{\mathbf{k}\mathbf{k}^{\prime}}$.
Therefore, we obtain the physical current dressed by Landau backflow,
\[
\mathcal{J}^{\omega}_{i}=ev^{0}_{i}+\gamma^{(\mathrm{int},\omega)}_{i}=(1+f\chi)ev^{\mathrm{qp}}_{i}
\]
with $v^{\mathrm{qp}}_{i}$ being the quasiparticle velocity. In the
static limit, $R^{q}=-\chi$. The Bethe-Salpeter equation becomes
$\mathcal{J}^{q}_{i}=\mathcal{J}^{\omega}_{i}-f\chi\mathcal{J}^{q}_{i}$,
giving $\mathcal{J}^{q}_{i}=ev^{\mathrm{qp}}_{i}$.

We then discuss the anomalous potential term with the bare vertex
$\gamma^{(\mathrm{B})}_{i}(\mathbf{k})=ie\omega a_{i}(\mathbf{k})$.
We factor out the explicit $ie\omega$ by defining $\Lambda^{(\mathrm{B})}_{i}=ie\omega\mathcal{A}_{i}(\mathbf{k})$
with the full vertex $\Lambda^{(\mathrm{B})}_{i}$ dressed by interactions.
The Bethe-Salpeter equation reduces to 
\begin{equation}
\mathcal{A}_{i}(\mathbf{k};i\omega,\mathbf{q})=a_{i}(\mathbf{k})+\int_{\mathbf{p}}I_{\mathbf{k}\mathbf{p}}R(\mathbf{p};i\omega,\mathbf{q})\mathcal{A}_{i}(\mathbf{p};i\omega,\mathbf{q}).\label{eq:A-response-BS}
\end{equation}
In the static limit $R^{q}=-\chi$, we have $\mathcal{A}^{q}_{i}=(1+f\chi)^{-1}a_{i}$.
In the dynamic limit, we have $\mathcal{A}^{\omega}_{i}=a_{i}$.

\section{AHE and Drude weight in the BL-FL theory}

\label{sec:kinetic-transport}

In this section, we derive the Drude weight and intrinsic anomalous
Hall response from the energy functional of the BL-FL theory.

\subsection{Response theory from the free energy }

We start with the free energy 
\[
F[n,\mathbf{A}]=\delta E_{\mathrm{BL\text{-}FL}}[n,\mathbf{A}]-TS[n],
\]
with the entropy 
\[
S[n]=-\sum_{\mathbf{k}}\left[n_{\mathbf{k}}\ln n_{\mathbf{k}}+(1-n_{\mathbf{k}})\ln(1-n_{\mathbf{k}})\right].
\]
We expand the distribution function around equilibrium $n_{\mathbf{k}}=n^{0}_{\mathbf{k}}+\delta n_{\mathbf{k}}$
and determine the equilibrium distribution $n^{0}_{\mathbf{k}}$
by minimizing the free energy $F[n,\mathbf{A}]$. This gives $n^{0}_{\mathbf{k}}=n_{F}(\epsilon^{\mathrm{qp}}_{\mathbf{k}}-\mu)$.
The fluctuation part of the free energy $\delta F=F[n,\mathbf{A}]-F[n^{0},0]$
is given by 
\begin{align}
\delta F= & \frac{1}{2}\int_{\mathbf{k}}\chi^{-1}_{\mathbf{k}}(\delta n_{\mathbf{k}})^{2}+\frac{1}{2}\int_{\mathbf{k}\mathbf{k}^{\prime}}f_{\mathbf{k}\mathbf{k}^{\prime}}\delta n_{\mathbf{k}}\delta n_{\mathbf{k}^{\prime}}\nonumber \\
 & +\int_{\mathbf{k}}[\mathbf{A}\cdot\bm{\mathcal{J}}^{\omega}(\mathbf{k})+e\mathbf{E}\cdot\mathbf{a}(\mathbf{k})]\delta n_{\mathbf{k}}+\frac{1}{2}\int_{\mathbf{k}\mathbf{k}^{\prime}}\mathbf{A}\cdot\mathbf{\Lambda}_{\mathbf{k}\mathbf{k}^{\prime}}\delta n_{\mathbf{k}}\delta n_{\mathbf{k}^{\prime}}.\label{eq:deltaF}
\end{align}
Here $\mathbf{a}(\mathbf{k})$ is the Berry connection of the quasiparticles
and $\bm{\mathcal{J}}^{\omega}(\mathbf{k})$ is the dynamic-limit
current, 
\[
\bm{\mathcal{J}}^{\omega}(\mathbf{k})=e\mathbf{v}^{0}(\mathbf{k})+e\int_{\mathbf{k}^{\prime}}(\nabla_{\mathbf{k}}+\nabla_{\mathbf{k}^{\prime}})f_{\mathbf{k}\mathbf{k}^{\prime}}n^{0}_{\mathbf{k^{\prime}}}.
\]
The term $\frac{1}{2}\int_{\mathbf{k}}\chi^{-1}_{\mathbf{k}}(\delta n_{\mathbf{k}})^{2}$
with $\chi_{\mathbf{k}}=\beta n^{0}_{\mathbf{k}}(1-n^{0}_{\mathbf{k}})$
arises from the entropy term. We minimize $\delta F$ with respect
to $\delta n_{\mathbf{k}}$. To linear order in the external fields,
we obtain the equation 
\begin{equation}
\chi^{-1}_{\mathbf{k}}\delta n_{\mathbf{k}}+\int_{\mathbf{k}^{\prime}}f_{\mathbf{k}\mathbf{k}^{\prime}}\delta n_{\mathbf{k}^{\prime}}=-\mathbf{A}\cdot\bm{\mathcal{J}}^{\omega}(\mathbf{k})-e\mathbf{E}\cdot\mathbf{a}(\mathbf{k}).
\end{equation}
The term $\mathbf{A}\cdot\mathbf{\Lambda}_{\mathbf{k}\mathbf{k}^{\prime}}$
is dropped at leading order. If $M_{\mathbf{kk}^{\prime}}$ denotes
the inverse of the quadratic kernel 
\begin{equation}
\int_{\mathbf{p}}[\chi^{-1}_{\mathbf{k}}\delta_{\mathbf{kp}}+f_{\mathbf{kp}}]M_{\mathbf{pk}^{\prime}}=\delta_{\mathbf{k}\mathbf{k}^{\prime}},\label{eq:Meq}
\end{equation}
we get the expression for $\delta n_{\mathbf{k}}$ in terms of the
gauge field 
\begin{equation}
\delta n_{\mathbf{k}}=-\int_{\mathbf{k}^{\prime}}M_{\mathbf{kk}^{\prime}}[\mathbf{A}\cdot\bm{\mathcal{J}}^{\omega}(\mathbf{k}^{\prime})+e\mathbf{E}\cdot\mathbf{a}(\mathbf{k}^{\prime})].
\end{equation}
Thus, we have the free energy to quadratic order 
\begin{equation}
\delta F=-\frac{1}{2}\int_{\mathbf{k}\mathbf{k}^{\prime}}s_{\mathbf{k}}M_{\mathbf{kk}^{\prime}}s_{\mathbf{k}^{\prime}}+\mathcal{O}(A^{3},A^{2}E,AE^{2}),\label{eq:sMs}
\end{equation}
where $s_{\mathbf{k}}=\mathbf{A}\cdot\bm{\mathcal{J}}^{\omega}(\mathbf{k})+e\mathbf{E}\cdot\mathbf{a}(\mathbf{k})$.
The $\mathbf{A}\cdot\mathbf{\Lambda}_{\mathbf{k}\mathbf{k}^{\prime}}$
term in Eq. (\ref{eq:deltaF}) does not modify the Drude weight or
the linear AHE response. It can contribute to the nonlinear response.

\emph{Drude weight.---}The $A_{i}A_{j}$ term of Eq.~\eqref{eq:sMs}
determines the Drude weight. In matrix form, the response kernel is
\[
M=(\chi^{-1}+f)^{-1}=\chi(1+f\chi)^{-1}
\]
 from Eq.~\eqref{eq:Meq}. The Drude weight can be obtained from
the derivatives of the free energy $D_{ij}=\frac{\delta^{2}F}{\delta A_{i}\delta A_{j}}$,
giving 
\begin{equation}
D_{ij}=e^{2}\int_{\mathbf{k}}\chi_{\mathbf{k}}v^{\mathrm{qp}}_{\mathbf{k},i}v^{\mathrm{qp}}_{\mathbf{k},j}+e^{2}\int_{\mathbf{k}\mathbf{k}'}\chi_{\mathbf{k}}v^{\mathrm{qp}}_{i}(\mathbf{k})\chi_{\mathbf{k}'}f_{\mathbf{k}\mathbf{k}'}v^{\mathrm{qp}}_{\mathbf{k}'j}.\label{eq:sm_Drude_freeenergy}
\end{equation}
For an isotropic Fermi liquid, $D(0)=D^{\mathrm{qp}}\left(1+\frac{F_{1}}{d}\right)$,
with $F_{1}$ being the $\ell=1$ Landau parameter. Here $D^{\mathrm{qp}}$
is the Drude weight associated with the quasiparticles without the
Landau backflow.

\emph{Anomalous Hall response.---}The mixed $A_{i}E_{j}$ term of
Eq.~\eqref{eq:sMs} determines the anomalous Hall conductivity $\sigma^{\mathrm{AHE}}_{ij}=\frac{\delta^{2}F}{\delta A_{i}\delta E_{j}}-\frac{\delta^{2}F}{\delta A_{j}\delta E_{i}}$.
Explicitly, we have 
\[
\sigma^{\mathrm{AHE}}_{ij}=e\left[\int_{\mathbf{k}\mathbf{k}'}\mathcal{J}^{\omega}_{i}(\mathbf{k})M_{\mathbf{kk}^{\prime}}a_{j}(\mathbf{k}^{\prime})-(i\leftrightarrow j)\right].
\]
The Landau-interaction dressing cancels completely between $\bm{\mathcal{J}}^{\omega}=(1+f\chi)e\mathbf{v}^{\mathrm{qp}}$
and $M=\chi(1+f\chi)^{-1}$. Therefore, 
\begin{align}
\sigma^{\mathrm{AHE}}_{ij} & =e^{2}\int_{\mathbf{k}}\chi_{\mathbf{k}}\left[v^{\mathrm{qp}}_{\mathbf{k},i}a_{j}(\mathbf{k})-v^{\mathrm{qp}}_{\mathbf{k},j}a_{i}(\mathbf{k})\right]\nonumber \\
 & =\frac{e^{2}}{\hbar}\int_{\mathbf{k}}n_{F}(\epsilon^{\mathrm{qp}}_{\mathbf{k}}-\mu)\,\Omega^{\mathrm{qp}}_{ij}(\mathbf{k}).
\end{align}
Here $\hbar$ has been restored in the second line. Thus, the anomalous
Hall conductivity contains no additional Landau backflow factor. Interactions
can still affect the Hall response indirectly through the renormalized
quasiparticle dispersion and Berry curvature. Including the fully
occupied remote bands, $\sigma^{\mathrm{AHE}}_{ij}=C_{\mathrm{filled}}\frac{e^{2}}{h}+\frac{e^{2}}{\hbar}\int_{\mathrm{active}}n_{F}(\epsilon^{\mathrm{qp}}_{\mathbf{k}}-\mu)\,\Omega^{\mathrm{qp}}_{ij}(\mathbf{k})$.

\section{Generic perturbations}

\subsection{General formula}

In this section, we discuss a generalized Peierls substitution for
a generic response. Consider a slow external force $X_{\alpha}(t)$
that deforms the quasiparticle momentum as 
\[
\mathbf{k}^{X}=\mathbf{k}+\sum_{\alpha}\mathbf{G}_{\alpha}(\mathbf{k})X_{\alpha},
\]
where $\alpha$ labels the components of the external source and $\mathbf{G}_{\alpha}(\mathbf{k})$
specifies how the source moves a state at momentum $\mathbf{k}$.
For a spatially uniform, slowly varying source, the instantaneous
Bloch state follows the corresponding generalized Houston trajectory,
$|u_{\mathbf{k}}(t)\rangle=|u_{\mathbf{k}(t)}\rangle$, where $\mathbf{k}(t)$
satisfies the equation of motion $\dot{\mathbf{k}}=\sum_{\alpha}\mathbf{G}_{\alpha}(\mathbf{k})\dot{X}_{\alpha}$.
Then we obtain the anomalous potential term 
\begin{equation}
L_{B}=\sum_{\mathbf{k}}\dot{\mathbf{k}}\cdot\mathbf{a}(\mathbf{k})f^{\dagger}_{\mathbf{k}}f_{\mathbf{k}}=\sum_{\mathbf{k}}\sum_{\alpha}\dot{X}_{\alpha}\mathbf{G}_{\alpha}(\mathbf{k})\cdot\mathbf{a}(\mathbf{k})f^{\dagger}_{\mathbf{k}}f_{\mathbf{k}}.
\end{equation}
On the other hand, perturbation theory gives the change of the Bloch
wave under the external force, 
\[
|\delta u_{\mathbf{k}}\rangle=\sum_{i\alpha}X_{\alpha}G_{i\alpha}(\mathbf{k})|\mathcal{D}_{i}u_{\mathbf{k}}\rangle,
\]
where $\mathcal{D}_{i}=\partial_{k_{i}}-ia_{i}(\mathbf{k})$ is the
covariant derivative. Given the projected density-density interaction
in the forward-scattering channel $H_{\mathrm{int}}\rightarrow f^{0}_{\mathbf{kk}^{\prime}}n_{\mathbf{k}}n_{\mathbf{k}^{\prime}}$,
with $f^{0}_{\mathbf{kk}^{\prime}}=-V_{\mathbf{k}-\mathbf{k}^{\prime}}|\langle u_{\mathbf{k}}|u_{\mathbf{k}^{\prime}}\rangle|^{2}$,
we have the interaction-induced current operator, 
\begin{equation}
J^{X}_{\alpha}(\mathbf{k})=-\frac{1}{V}\sum_{\mathbf{k}^{\prime}}n_{\mathbf{k}}n_{\mathbf{k}^{\prime}}\sum_{i}\left[G_{i\alpha}(\mathbf{k})\partial_{k_{i}}+G_{i\alpha}(\mathbf{k}^{\prime})\partial_{k^{\prime}_{i}}\right]f^{0}_{\mathbf{kk}^{\prime}}.\label{eq:Jx}
\end{equation}
Within the BL-FL theory, this current contributes the following term
to the energy functional 
\begin{equation}
\delta E[n,X]=\frac{1}{V}\sum_{\mathbf{k}\mathbf{k}^{\prime}}n_{\mathbf{k}}n_{\mathbf{k}^{\prime}}\sum_{i\alpha}X_{\alpha}\left[G_{i\alpha}(\mathbf{k})\partial_{k_{i}}+G_{i\alpha}(\mathbf{k}^{\prime})\partial_{k^{\prime}_{i}}\right]f_{\mathbf{kk}^{\prime}}.
\end{equation}
Consider a vector potential $\mathbf{A}^{\lambda}$ that shifts momentum
according to $\mathbf{k}\rightarrow\mathbf{k}+\lambda_{\mathbf{k}}\mathbf{A}^{\lambda}$,
that is, $G_{i\alpha}(\mathbf{k})=\lambda_{\mathbf{k}}\delta_{\alpha i}$.
The current operator in Eq.~\eqref{eq:Jx} becomes 
\begin{equation}
J^{\lambda}_{i}(\mathbf{k})=-\frac{1}{V}\sum_{\mathbf{k}^{\prime}}n_{\mathbf{k}}n_{\mathbf{k}^{\prime}}\left[\lambda_{\mathbf{k}}\partial_{k_{i}}+\lambda_{\mathbf{k}^{\prime}}\partial_{k^{\prime}_{i}}\right]f^{0}_{\mathbf{kk}^{\prime}},
\end{equation}
and the corresponding term in the BL-FL energy functional is 
\begin{equation}
\delta E[n,\lambda]=\frac{1}{V}\sum_{\mathbf{k}\mathbf{k}^{\prime}}n_{\mathbf{k}}n_{\mathbf{k}^{\prime}}\mathbf{A}^{\lambda}\cdot\left[\lambda_{\mathbf{k}}\nabla_{\mathbf{k}}+\lambda_{\mathbf{k}^{\prime}}\nabla_{\mathbf{k}^{\prime}}\right]f_{\mathbf{kk}^{\prime}}.
\end{equation}

\subsubsection{Thermal current}

For the thermal current, Luttinger's gravitational vector potential
$\mathbf{A}^{g}$ produces the low-energy shift $\mathbf{k}\rightarrow\mathbf{k}+\xi_{\mathbf{k}}\mathbf{A}^{g}$
with $\xi_{\mathbf{k}}=\epsilon^{\mathrm{qp}}_{\mathbf{k}}-\mu$ and
$\lambda_{\mathbf{k}}=\xi_{\mathbf{k}}$. In the BL-FL theory, the interaction-induced
heat current is 
\begin{align}
\mathcal{J}^{Q,\mathrm{int}}_{\alpha}(\mathbf{k}) & =\frac{1}{V}\sum_{\mathbf{k}^{\prime}}n_{\mathbf{k}^{\prime}}\left[\xi_{\mathbf{k}}\partial_{k_{\alpha}}+\xi_{\mathbf{k^{\prime}}}\partial_{k^{\prime}_{\alpha}}\right]f_{\mathbf{kk}^{\prime}}\\
 & =\frac{1}{V}\sum_{\mathbf{k}^{\prime}}n_{\mathbf{k}^{\prime}}\left[\frac{\xi_{\mathbf{k}}+\xi_{\mathbf{k}^{\prime}}}{2}\left(\partial_{k_{\alpha}}+\partial_{k^{\prime}_{\alpha}}\right)+\frac{\xi_{\mathbf{k}}-\xi_{\mathbf{k^{\prime}}}}{2}\left(\partial_{k_{\alpha}}-\partial_{k^{\prime}_{\alpha}}\right)\right]f_{\mathbf{kk}^{\prime}}.
\end{align}
For a slowly varying gravitational vector potential, $\mathbf{E}^{g}=-\partial_{t}\mathbf{A}^{g}=-\frac{\nabla T}{T}$,
and the crystal momentum $\mathbf{k}$ follows the equation of motion $\dot{\mathbf{k}}=-\xi_{\mathbf{k}}\frac{\nabla T}{T}$.
The corresponding anomalous potential is 
\[
\delta L_{B}=\frac{\xi_{\mathbf{k}}}{T}\nabla T\cdot\mathbf{a}(\mathbf{k}).
\]
This gives the thermal analogue of the Wannier-Stark energy contributed
by the Berry connection.

\subsubsection{Frame rotation}

A rotation of the local frame by an angle $\theta(t)$ about $\hat{z}$
acts on momentum as 
\begin{equation}
\mathbf{k}_{\theta}=R^{-1}_{\theta}\mathbf{k}=\mathbf{k}-\theta\hat{z}\times\mathbf{k}+\mathcal{O}(\theta^{2}).
\end{equation}
Thus, $\dot{\mathbf{k}}=-\omega_{z}\hat{z}\times\mathbf{k}$ with
$\omega_{z}=\dot{\theta}$. If the projected basis also transforms
under an internal rotation $|u_{\mathbf{k}};\theta\rangle=e^{-i\theta S^{\mathrm{orbital}}_{z}}|u_{R^{-1}_{\theta}\mathbf{k}}\rangle$,
the the spatial rotation generates an additional internal spin-connection
term. Thus, we have the anomalous potential term 
\begin{equation}
\delta L_{B}=-\omega_{z}\left(\hat{z}\times\mathbf{k}\right)\cdot\mathbf{a}(\mathbf{k})-\omega_{z}s_{z}(\mathbf{k})
\end{equation}
with the projected spin connection $s_{z}(\mathbf{k})=\langle u_{\mathbf{k}}|S^{\mathrm{orbital}}_{z}|u_{\mathbf{k}}\rangle$.
In the interaction, the internal unitary rotation cancels in the form factor  $ \langle u_{\mathbf k};\theta |u_{\mathbf k};\theta \rangle = \langle u_{R^{-1}_\theta\mathbf k}|u_{R^{-1}_\theta\mathbf k}\rangle$. Thus, the interaction-induced rotation current operator is
\begin{equation}
J^{\theta}_{z}(\mathbf{k})=-\frac{1}{V}\sum_{\mathbf{k}^{\prime}}n_{\mathbf{k}}n_{\mathbf{k}^{\prime}}\left[(\hat{z}\times\mathbf{k})\cdot\nabla_{\mathbf{k}}+(\hat{z}\times\mathbf{k}^{\prime})\cdot\nabla_{\mathbf{k}^{\prime}}\right]f^{0}_{\mathbf{kk}^{\prime}}.
\end{equation}

\subsubsection{Vielbein}

We start with the perturbation from the vielbein field to consider
the forces due to strain, shear, and rotation. We take a time-dependent
spatial vielbein $e^{a}_{i}(t)$ and its inverse $e^{i}_{a}(t)$:
$e^{i}_{a}e^{b}_{i}=\delta^{b}_{a}$. Let $q_{i}$ be the conserved
coordinate momentum; then the momentum entering the local Bloch Hamiltonian
is $k_{a}(t)=e^{i}_{a}q_{i}$. From $\dot{k}_{a}=\dot{e}^{i}_{a}q_{i}$
and $\dot{e}^{i}_{a}=-e^{j}_{a}\dot{e}^{b}_{j}e^{i}_{b}$, we obtain
\begin{equation}
\dot{k}_{a}=-\Gamma_{ab}k_{b},\quad\Gamma_{ab}\equiv e^{i}_{a}\dot{e}^{b}_{i}.
\end{equation}
For a weak deformation of the undeformed frame, $e^{a}_{i}\simeq\delta^{a}_{i}$,
and we simply have $\dot{k}_{i}=-\Gamma_{ij}k_{j}$. Hence, we
have the vielbein-induced Berry term 
\begin{equation}
\delta L_{B}=-\sum_{\mathbf{k}}\Gamma_{ij}k_{j}a_{i}(\mathbf{k})f^{\dagger}_{\mathbf{k}}f_{\mathbf{k}}.
\end{equation}

We can decompose $\Gamma_{ij}$ into symmetric and antisymmetric parts:
$\Gamma_{ij}=\Gamma_{(ij)}+\Gamma_{[ij]}$ with $\Gamma_{(ij)}=\frac{1}{2}(\Gamma_{ij}+\Gamma_{ji})$
and $\Gamma_{[ij]}=\frac{1}{2}(\Gamma_{ij}-\Gamma_{ji})$. The symmetric
part describes the strain and shear $u_{ij}$ with the relation $\Gamma_{(ij)}=\dot{u}_{ij}$,
giving $\dot{k}_{i}=-\dot{u}_{ij}k_{j}$. The corresponding term in
BL-FL is 
\begin{equation}
\delta E[n,u]=\frac{1}{V}\sum_{\mathbf{k}\mathbf{k}^{\prime}}n_{\mathbf{k}}n_{\mathbf{k}^{\prime}}\sum_{ij}u_{ij}(k_{j}\partial_{k_{i}}+k_{i}\partial_{k_{j}})f_{\mathbf{kk}^{\prime}}.
\end{equation}
Rotation is described by the antisymmetric part $\Gamma_{[ij]}=\omega_{z}\epsilon_{ij}$.

\section{Numerics on the Wilson-Dirac model}

\label{sec:hf_model}

We give the numerical details of the Wilson-Dirac model and the self-consistent
Hartree-Fock calculation used to test the BL-FL theory.

\subsection{Wilson-Dirac model}

We consider spinless fermions with two orbitals per unit cell, $H_{0}=\sum_{\mathbf{k}}c^{\dagger}_{\mathbf{k}}H_{0}(\mathbf{k})c_{\mathbf{k}},\qquad c_{\mathbf{k}}=\begin{pmatrix}c_{\mathbf{k}1}\\
c_{\mathbf{k}2}
\end{pmatrix}.$ The Bloch projectors are obtained from the vector $\mathbf{d}(\mathbf{k})$ with components
\begin{equation}
d_{x}(\mathbf{k})=v\sin k_{x},\quad d_{y}(\mathbf{k})=v\sin k_{y},\quad d_{z}(\mathbf{k})=M-2B(2-\cos k_{x}-\cos k_{y}),
\end{equation}
and $\hat{\mathbf{d}}(\mathbf{k})=\frac{\mathbf{d}(\mathbf{k})}{|\mathbf{d}(\mathbf{k})|}.$
For $0<M<4B$, the two bands carry opposite nonzero Chern numbers.
To control the band dispersion and the remote-band gap independently,
we use a deformed Wilson-Dirac Hamiltonian 
\begin{equation}
H_{0}(\mathbf{k})=s\zeta(\mathbf{k})\sigma_{0}+\Delta\hat{\mathbf{d}}(\mathbf{k})\cdot\bm{\sigma},\label{eq:sm_wilson_dirac}
\end{equation}
with $\zeta(\mathbf{k})=|\mathbf{d}(\mathbf{k})|.$ The eigenvalues
of $H_{0}$ are $\epsilon^{0}_{\pm}(\mathbf{k})=s\zeta(\mathbf{k})\pm\Delta.$
Here $s$ controls the bandwidth, while $2\Delta$ is the band gap.
Thus, we can tune the band gap and bandwidth while keeping the quantum
geometry fixed. We consider a screened density-density interaction 
\begin{equation}
H_{\mathrm{int}}=\frac{1}{2N}\sum_{\mathbf{q}}V(\mathbf{q})\rho_{\mathbf{q}}\rho_{-\mathbf{q}},
\end{equation}
where $\rho_{\mathbf{q}}=\sum_{\mathbf{k},\alpha}c^{\dagger}_{\mathbf{k}+\mathbf{q},\alpha}c_{\mathbf{k},\alpha}$
and $V(\mathbf{q})=\frac{g}{1+2\ell^{2}_{s}(2-\cos q_{x}-\cos q_{y})}$.
We take the lower band to be fully filled and tune the chemical potential
through the upper band. After projection onto the active band, the
HF energy of the upper band, $\epsilon^{0}_{\mathbf{k}}\equiv\epsilon^{0}_{+}(\mathbf{k})$, is 
\begin{equation}
\epsilon^{\mathrm{HF}}_{\mathbf{k}}=\epsilon^{0}_{\mathbf{k}}+\frac{1}{N}\sum_{\mathbf{k}'}f^{\mathrm{HF}}_{\mathbf{k}\mathbf{k}'}n^{\mathrm{HF}}_{\mathbf{k}'},\label{eq:sm_projected_hf}
\end{equation}
where $n^{\mathrm{HF}}_{\mathbf{k}}=n_{F}\!\left(\epsilon^{\mathrm{HF}}_{\mathbf{k}}-\mu\right)$.
The Landau function in the HF approximation is 
\begin{equation}
f^{\mathrm{HF}}_{\mathbf{k}\mathbf{k}'}=V(0)-V(\mathbf{k}-\mathbf{k}')\operatorname{Tr}\left[P^{0}_{\mathbf{k}}P^{0}_{\mathbf{k}'}\right].\label{eq:sm_hf_landau_function}
\end{equation}
The interaction contribution to the conserved HF current is 
\begin{align}
\gamma^{\mathrm{(int),HF}}_{i}(\mathbf{k}) & =\frac{e}{N}\sum_{\mathbf{k}'}n^{\mathrm{HF}}_{\mathbf{k}'}\left(\partial_{k_{i}}+\partial_{k_{i}'}\right)f^{\mathrm{HF}}_{\mathbf{k}\mathbf{k}'}\label{eq:sm_hf_qg_current}\\
 & =-\frac{e}{N}\sum_{\mathbf{k}'}n^{\mathrm{HF}}_{\mathbf{k}'}V(\mathbf{k}-\mathbf{k}')\left(\partial_{k_{i}}+\partial_{k_{i}'}\right)\operatorname{Tr}\left[P^{0}_{\mathbf{k}}P^{0}_{\mathbf{k}'}\right].
\end{align}
Eqs. $\eqref{eq:sm_projected_hf}$ and $\eqref{eq:sm_hf_landau_function}$
are iterated until the change of the HF energy is below a fixed numerical
tolerance.

\subsection{Drude weight }

\label{sec:hf_twist_protocol}

The Drude weight is evaluated in two independent ways. First, we apply
a uniform vector potential $A_{x}$. The bare dispersion and projected
interaction kernel are both evaluated at the shifted Bloch momenta.
The corresponding HF energy per unit cell is 
\begin{equation}
E_{\mathrm{HF}}(A_{x})=\frac{1}{N}\sum_{\mathbf{k}}n^{\mathrm{HF}}_{\mathbf{k}}\epsilon^{0}_{\mathbf{k}}(A_{x})+\frac{1}{2N}\sum_{\mathbf{k}}n^{\mathrm{HF}}_{\mathbf{k}}\Sigma^{\mathrm{HF}}_{\mathbf{k}}(A_{x}),
\end{equation}
with $\Sigma^{\mathrm{HF}}_{\mathbf{k}}(A_{x})=\frac{1}{N}\sum_{\mathbf{k}'}f^{\mathrm{HF}}_{\mathbf{k}\mathbf{k}'}(A_{x})n^{\mathrm{HF}}_{\mathbf{k}'}$,
and the Kohn stiffness is 
\begin{equation}
D^{\mathrm{Kohn}}_{xx}=\pi\left.\frac{\partial^{2}F_{\mathrm{HF}}(A_{x})}{\partial A^{2}_{x}}\right|_{A_{x}=0},\label{eq:sm_kohn_drude}
\end{equation}
with the free energy $F_{\mathrm{HF}}(A_{x})=E_{\mathrm{HF}}(A_{x})-TS(A_{x})$.
Numerically, the curvature is extracted with a fourth-order symmetric
five-point stencil. Second, we evaluate the finite-temperature Landau
response. With $\chi_{\mathbf{k}}=-\frac{\partial n_{F}}{\partial\epsilon^{\mathrm{HF}}_{\mathbf{k}}}$
and $v^{\mathrm{HF}}_{x,\mathbf{k}}=\partial_{k_{x}}\epsilon^{\mathrm{HF}}_{\mathbf{k}}$,
the dressed velocity is 
\[
\mathcal{V}_{x,\mathbf{k}}=v^{\mathrm{HF}}_{x,\mathbf{k}}+\frac{1}{N}\sum_{\mathbf{k}'}f^{\mathrm{HF}}_{\mathbf{k}\mathbf{k}'}\chi_{\mathbf{k}'}v^{\mathrm{HF}}_{x,\mathbf{k}'}.
\]
The Drude weight is then 
\begin{equation}
D_{xx}=\frac{\pi e^{2}}{N}\sum_{\mathbf{k}}\chi_{\mathbf{k}}v^{\mathrm{HF}}_{x,\mathbf{k}}\mathcal{V}_{x,\mathbf{k}}.\label{eq:sm_landau_drude}
\end{equation}
Agreement between Eqs. $\eqref{eq:sm_kohn_drude}$ and $\eqref{eq:sm_landau_drude}$
provides the numerical test shown in Fig. 1(c) of the main text.

\end{document}